\documentclass[aps,prb,twocolumn,showpacs,10pt,floatfix,longbibliography]{revtex4-1}

\usepackage[pdftex]{graphicx} 
\usepackage{epstopdf}
\usepackage{verbatim}
\usepackage{color}
\usepackage{subfigure}
\usepackage{tabularx}
\usepackage{amsfonts}
\usepackage{wasysym}
\usepackage{bm} 
\usepackage[hypertexnames=false]{hyperref}

\newcommand{\be}{\begin{equation}}
\newcommand{\bea}{\begin{eqnarray}}
\newcommand{\ee}{\end{equation}}
\newcommand{\eea}{\end{eqnarray}}

\newcommand{\vev}[1]{\left\langle#1\right\rangle}

\newcommand{\rb}{\right)}
\newcommand{\lb}{\left(}

\begin{document}
\title{An anisotropic spin model of strong spin-orbit-coupled triangular antiferromagnets}
\author{Yao-Dong Li$^1$}
\author{Xiaoqun Wang$^{2,3}$}
\author{Gang Chen$^{4,5}$}
\email{gchen$_$physics@fudan.edu.cn}
\affiliation{${}^1$School of Computer Science, Fudan University, 
Shanghai, 200433, People's Republic of China}
\affiliation{${}^2$Department of Physics, 
Renmin University of China, Beijing 100872, People's Republic of China}
\affiliation{${}^3$Department of Physics and astronomy, 
Innovative Center for Advanced Microstructures, 
Shanghai Jiao Tong University, Shanghai 200240, People's Republic of China}
\affiliation{${}^4$State Key Laboratory of Surface Physics, 
Center for Field Theory and Particle Physics, 
Department of Physics, Fudan University, Shanghai 200433, People's Republic of China}
\affiliation{${}^5$Collaborative Innovation Center of Advanced Microstructures,
Fudan University, Shanghai, 200433, People's Republic of China}

\begin{abstract}
Motivated by the recent experimental progress on the 
strong spin-orbit-coupled rare earth triangular antiferromagnet, 
we analyze the highly anisotropic spin model that describes the interaction 
between the spin-orbit-entangled Kramers' doublet local moments 
on the triangular lattice. We apply the Luttinger-Tisza method, 
the classical Monte Carlo simulation, and the self-consistent 
spin wave theory to analyze the anisotropic spin Hamiltonian. 
The classical phase diagram includes the 120$^\circ$ state and two distinct 
stripe ordered phases. The frustration is very strong and significantly 
suppresses the ordering temperature in the regimes close to the phase boundary 
between two ordered phases. Going beyond the semiclassical analysis, 
we include the quantum fluctuations of the spin moments 
within a self-consistent Dyson-Maleev spin-wave treatment.
We find that the strong quantum fluctuations melt the magnetic order
in the frustrated regions. We explore the magnetic excitations 
in the three different ordered phases as well as in strong magnetic fields. 
Our results provide a guidance for the future theoretical study of the generic model 
and are broadly relevant for strong spin-orbit-coupled triangular antiferromagnets 
such as YbMgGaO$_4$, RCd$_3$P$_3$, RZn$_3$P$_3$, RCd$_3$As$_3$, RZn$_3$As$_3$,
and R$_2$O$_2$CO$_3$. 
\end{abstract}

\date{\today}

\maketitle

\section{Introduction}
\label{sec1}

Since the discovery of topological insulator~\cite{KaneReview}, 
spin-orbit coupling (SOC) has become one of the central topics 
in modern condensed matter physics.  
While topological insulator is the band structure topological property of 
non-interacting electrons, the interplay of 
strong spin-orbit coupling and strong electron correlation 
is one of the central questions in the field of 
strong correlation physics~\cite{WCKB}. 
In the recent years, there have been 
intense interests and activities in the heavy-element based 
materials where both strong spin-orbit coupling and 
strong electron correlations are present. 
The spin-orbit entanglement in strongly correlated electron systems 
can give rise to unprecedented and realistic models that may support 
novel phases and phenomena. 

Magnets with rare earth elements are natural physical systems 
to search for strong correlation physics with strong SOC. 
In the rare earth magnets, the correlation is often quite strong
and the $4f$ electrons are very localized. The atomic spin-orbit coupling
entangles the spin and orbital angular momenta and leads to 
a spin-orbit-entangled local moments. Recently, a Ytterbium
based rare earth magnet, YbMgGaO$_4$, has been synthesized and characterized~\cite{Yuesheng2015,Yueshengscirep2015}. 
The magnetic ions, Yb$^{3+}$, form a perfect triangular lattice. 
The SOC and the crystal electric field together lead to a Kramers'
doublet for the Yb$^{3+}$ ion. This Kramers' doublet is described 
by an effective spin-1/2 local moment. The thermodynamic and NMR measurements 
found that the system remains disordered down to 60mK~\cite{Yuesheng2015}. 
More recently, another rare-earth triangular antiferromagnet CeCd$_3$P$_3$ 
was studied experimentally~\cite{1604}. 
Although this material remains paramagnetic 
down to 0.48K and this temperature is probably 
not very low by the $4f$ electrons' standard,
as we show in Table~\ref{tab1}, CeCd$_3$P$_3$ 
and the rare-earth oxy-carbonates R$_2$O$_2$CO$_3$
represent new family of rare-earth 
triangular antiferromagnets that need further investigation~\cite{1604,Nientiedt1999,Yamada2010,Stoyko2011}.
Like the Yb$^{3+}$ ion in YbMgGaO$_4$, the Ce$^{3+}$ ion 
in CeCd$_3$P$_3$ experiences the same $D_{3d}$ crystal field 
and is also described by an effective spin-1/2 Kramers 
doublet~\cite{1604}. Partly motivated by these experiments, 
in this paper we consider the generic spin model that generally 
describes the spin-orbit-entangled Kramers' doublets on the triangular 
lattice and study the magnetic phase diagram and the magnetic 
excitation of this new model. 

\begin{figure}[t]
{\includegraphics[width=8cm]{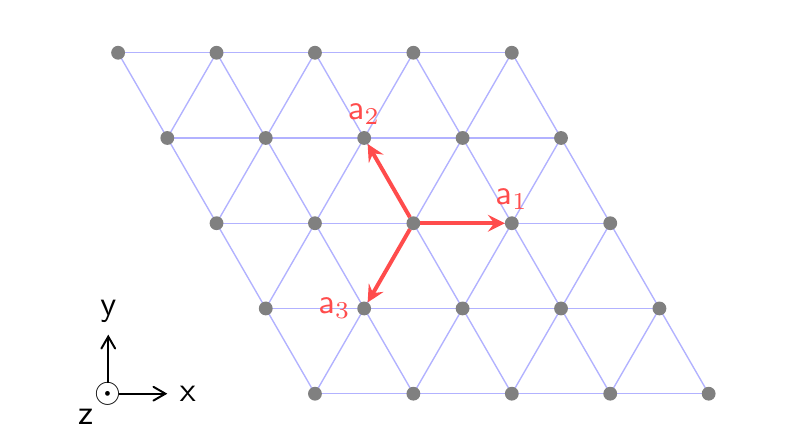}}
 \caption{(Color online.)
    Triangular lattice and the three nearest neighbors. 
    The inset defines the coordinate system for the spin components. }
	\label{fig1}
\end{figure}

\begin{table*}[t]
\centering
\begin{tabular}{lccccccc}
\hline\hline \\[0.1ex]
Compound & Magnetic ion & Space group & Local moment & $\Theta_{\text{CW}}$ (K) & Magnetic transition & Frustration para. $f$  & Ref
\\[1ex]
YbMgGaO$_4$ & Yb$^{3+}$ (4$f^{13}$) & R$\bar{3}$m & Kramers doublet &$-4$ & PM down to 60mK & $f>66$&
\onlinecite{Yueshengscirep2015}
\\[1ex]
CeCd$_3$P$_3$ & Ce$^{3+}$ (4$f^1$) & P$6_3/mmc$ & Kramers doublet & $-60$ & PM down to 0.48K & $f>200$ & 
\onlinecite{1604}
\\[1ex]
CeZn$_3$P$_3$ & Ce$^{3+}$ (4$f^1$) & P$6_3/mmc$ & Kramers doublet & $-6.6$ & AFM order at 0.8K & $f=8.2$ & \onlinecite{Yamada2010}
\\[1ex]
CeZn$_3$As$_3$ & Ce$^{3+}$ (4$f^1$) & P$6_3/mmc$ & Kramers doublet & $-62$ &  unknown & unknown &
\onlinecite{Stoyko2011}
\\[1ex]
PrZn$_3$As$_3$ & Pr$^{3+}$ (4$f^{2}$) & P$6_3/mmc$ & Non-Kramers doublet & $-18$ & unknown & unknown &\onlinecite{Stoyko2011}
\\[1ex]
NdZn$_3$As$_3$ & Nd$^{3+}$ (4$f^{3}$) & P$6_3/mmc$ & Kramers doublet &  $-11$ & unknown & unknown
& \onlinecite{Stoyko2011}
\\[1ex]
Nd$_2$O$_2$CO$_3$ & Nd$^{3+}$ (4$f^{3}$) & P$6_3/mmc$ & Kramers doublet &  $-21.7$ & AFM order at 1.25K & $f= 17.4$ & \onlinecite{Arjun2016} 
\\[1ex]
Sm$_2$O$_2$CO$_3$ & Sm$^{3+}$ (4$f^{5}$) & P$6_3/mmc$ & Kramers doublet &  $-18$ & AFM order at 0.61K & $f= 31$ & \onlinecite{Arjun2016} 
\\[1ex]
Dy$_2$O$_2$CO$_3$ & Dy$^{3+}$ (4$f^{9}$) & P$6_3/mmc$ & Kramers doublet &  $-10.6$ & AFM order at 1.21K & $f= 8.8$ & \onlinecite{Arjun2016} 
\\[1ex]
\hline\hline
\end{tabular}
\caption{A list of rare-earth triangular antiferromagnets. Note the Curie-Weiss temperatures ($\Theta_{\text{CW}}$) for the second to the sixth compounds are obtained from the 
magnetic susceptibility measurments above 50K. 
Here, `PM' refers to paramagnetic and `AFM' refers to antiferromagnetic.
The frustration parameter $f$ is defined in Sec.~\ref{sec3b}.}
\label{tab1}
\end{table*}

Due to the spin-orbit-entangled nature of the Kramers' doublets,   
the interaction between the effective spin-1/2 moments is 
{\sl anisotropic} both in the effective spin space and in the position space~\cite{Chen2008,WCKB,JXLi2015,Chen2010,Curnoe2008,Onoda2010,Chen2011,Jackeli2009}.
Therefore, the spin interaction depends on the bond orientations. 
This is one of the key propeties of the strong spin-orbit-coupled magnets. 
The most generic spin Hamiltonian allowed by the space group symmetry of 
the rare earth triangular system is given by~\cite{Yuesheng2015}
\begin{eqnarray}
	{\mathcal H} &=& \sum_{\vev{ij}}   
	J_{zz} S_i^z S_j^z + J_{\pm} ( S_i^+ S_j^- + S_i^- S_j^+ ) 
	\nonumber \\
	&& 
	+ J_{\pm\pm} (  \gamma_{ij} S_i^+ S_j^+ +  \gamma_{ij}^{\ast} S_i^- S_j^- ) 
	\nonumber \\
    && 
    - \frac{i J_{z\pm}}{2} 
    \big[ (\gamma_{ij}^{\ast} S_i^+ - \gamma_{ij} S_i^- ) S_j^z 
    \nonumber \\
    && \quad\quad\quad
     +  S_i^z  (\gamma_{ij}^{\ast} S_j^+ - \gamma_{ij} S_j^- )  \big],
\label{eq1}
\end{eqnarray}
where $S_i^\pm = S_i^x \pm i S_i^y$, and $\gamma_{ij} =
\gamma_{ji} = 1, e^{i2\pi/3}, e^{-i2\pi/3}$ are the phase factors 
for the bond $ij$ along the \textbf{a$_1$}, \textbf{a$_2$}, 
\textbf{a$_3$} directions, respectively (see Fig.~\ref{fig1}). 
The first line of Eq.~(\ref{eq1}) is the standard XXZ model and is invariant
under the global spin rotation around the $z$ direction. 
Here we have chosen the coordinate system for the spin components 
to be identical with the one for the position space (see Fig.~\ref{fig1}). 
The $J_{\pm\pm}$ and $J_{z\pm}$ terms of Eq.~(\ref{eq1}) 
define the anisotropic interactions that arise naturally 
from the strong SOC.  

To study the generic spin model, we first carry out the semiclassical 
analysis of the generic spin Hamiltonian in Sec.~\ref{sec3}. 
Using the combined Luttinger-Tisza method and classical Monte Carlo simulation,
we first determine the classical ground state
phase diagram of the model. We find that 
the anisotropic $J_{\pm\pm}$ and $J_{z\pm}$ interactions 
compete with the XXZ part of the model and 
drive the system into two distinct stripe ordered phases. 
Then we implement the classical Monte Carlo 
simulation to uncover the classical 
magnetic orders at low temperatures. 
The ordering temperatures of different phases      
are determined as well. We find that the ordering temperatures  
are strongly suppressed near the 
the phase boundary between different ordered phases, 
suggesting the strong frustration in these regions. 

The existing experiments in YbMgGaO$_4$ suggest a disordered quantum ground state.
Our generic spin model is expected to describe the interaction between Yb$^{3+}$
local moments. Therefore, it is of importance to understand whether
the generic model may support a disordered ground state in the quantum regime, 
and which parameter regime such a disordered ground state may exist. 
For this purpose, in Sec.~\ref{sec4} we study the quantum fluctuation 
through a self-consistent Dyson-Maleev spin wave 
analysis and find that the quantum fluctuation 
is very strong and could melt the magnetic order 
in the parameter regimes near the phase boundary. 
We thus expect these regions may turn into a disordered
ground state when the quantum nature of the spins is considered. 

Since the generic spin model applies broadly to any other 
triangular system with Kramers' doublet and the 
long-range order should survive deep inside the 
ordered regions even for the quantum spins, 
these magnetic orders should be relevant for 
other triangular lattice magnets with strong SOC,
such as the RCd$_3$P$_3$, RZn$_3$P$_3$, RCd$_3$As$_3$, RZn$_3$As$_3$ family,
where R is a rare-earth element. 
It is likely that the magnetic order may appear in some of these materials.
In Sec.~\ref{sec5}, we compute the spin wave excitation in different ordered phases. 
Moreover, because the energy scale of the exchange coupling for the rare earth 
triangular magnets is usually very small, it is ready to apply
strong magnetic fields to fully polarize the spin moments. This 
allows a direct comparison between the theoretical results and 
the inelastic neutron scattering measurements in the future experiments 
both in YbMgGaO$_4$ and other relevant materials. 

The remainder of the paper is organized as follows.
In Sec.~\ref{sec2}, we explain the symmetry operation 
on the spin-orbit-entangled local moments and derive 
the generic spin model for the rare-earth triangular 
systems. In Sec.~\ref{sec3}, we carry out both
Luttinger-Tisza analysis and classical Monte
Carlo simulation, and determine the classical phase diagram. 
In Sec.~\ref{sec4}, we implement the self-consistent 
Dyson-Maleev spin wave calculation to study the quantum 
fluctuation in different ordered phase.
In Sec.~\ref{sec5}, we compute the spin-wave excitation
in the presence and absence of magnetic fields.
Finally in Sec.~\ref{sec6}, we discuss the connection with
the experiments and future theoretical directions.

\section{The generic spin Hamiltonian for Kramers' doublet}
\label{sec2}

We start with the symmetry transformation properties of the 
Kramers' doublet. While the discussion in this section is about  
the Yb$^{3+}$ ion in YbMgGaO$_4$, the symmetry analysis 
applies generally to any other Kramers' doublet 
that shares the same symmetry properties on the triangular lattice. 

\begin{figure}[t]
 {\includegraphics[width=.4\textwidth]{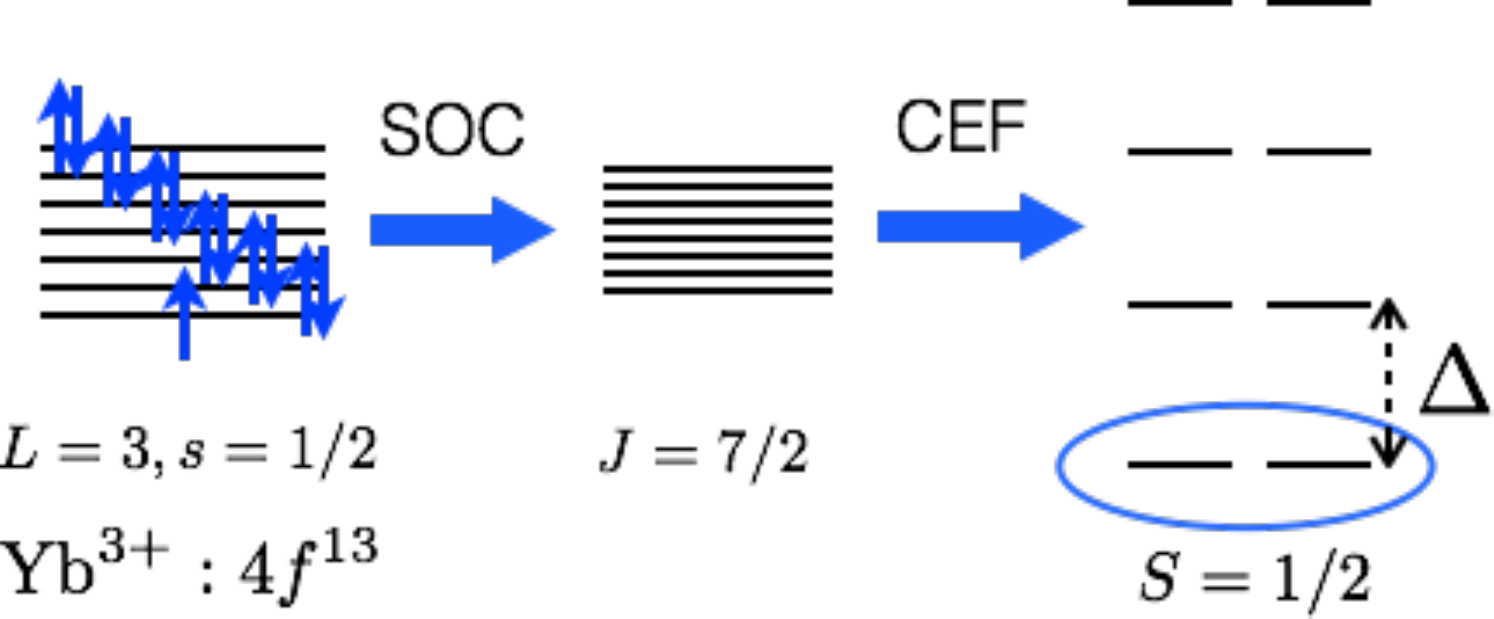}
 }
\caption{The formation of the local ground state Kramers' doublet
under the combination of spin-orbit coupling (SOC)
and the crystal electric field (CEF). Please refer the text 
for the detailed description.  
}
\label{fig2}
\end{figure}

The Yb$^{3+}$ ion contains thirteen $4f$ electrons. 
According to the Hund's rule, we should have the total spin 
$s=1/2$ and the orbital angular momentum $L=3$ for the Yb$^{3+}$ ion. 
The fourteen-fold spin and orbital degeneracy is lifted when 
the atomic SOC and the crystal electric field
are considered. For the $4f$ electrons, the atomic SOC 
should be considered before the crystal electric field.
As we show in Fig.~\ref{fig2}, the atomic SOC entangles
the orbital angular momentum and the total spin, leading to 
a total angular momentum $J=7/2$ with eight fold degeneracy. 
Just like the Yb$^{3+}$ ion for the pyrochlore ice material 
Yb$_2$Ti$_2$O$_7$~\cite{Ross2011}, the crystal electric field 
of the D$_{3d}$ point group further splits the eight $J=7/2$ 
states into four pairs of Kramers' doublets.  
The ground state doublet is well separated from other excited 
doublets with an energy gap $\Delta\sim 420$K and 
thus can be treated as an effective spin-1/2 degree of freedom at the 
temperature that is much lower than the energy gap~\cite{Yueshengscirep2015,Ross2011}. 
We introduce an effective spin-1/2 local moment, 
${\bf S}_i$, that operates on the local ground state Kramers' doublet. 
This effective spin-1/2 degree of freedom for the 
Yb$^{3+}$ ion is well supported by the low temperature magnetic entropy 
that is measured to be R$\ln 2$ per spin~\cite{Yueshengscirep2015,Yuesheng2015}. 

This effective spin, $\bf{S}$, results from the spin-orbit 
entanglement of the Yb$^{3+}$ $4f$ electrons. 
As a consequence, both the position and 
the orientation of the spins are transformed together 
under the space group symmetry operation, and the 
transformation is given as 
\begin{eqnarray}
{\bf S}_{\bf r} \rightarrow \text{Det}[\hat{O}] \cdot 
\hat{O}^{-1} \cdot {\bf S}_{\hat{O}\cdot {\bf r} + {\bf t} },
\end{eqnarray}
where $\hat{O}$ and ${\bf t}$ are the matrix and the vector
that specify the rotation part and the translation part of the 
space group operation, respectively. In contrast, 
in a magnetic system whose local moment is 
purely given by the total spin, the spin rotational symmetry 
would be decoupled from the space group symmetry operation. The latter 
merely acts on the positions of the spin moments and does not rotate
the spin components. This is the key difference 
between the strong spin-orbit coupled Mott insulators and 
a conventional Mott insulator with quenched orbital degrees of 
freedom. 

\begin{figure}[t]
{\includegraphics[width=.22\textwidth]{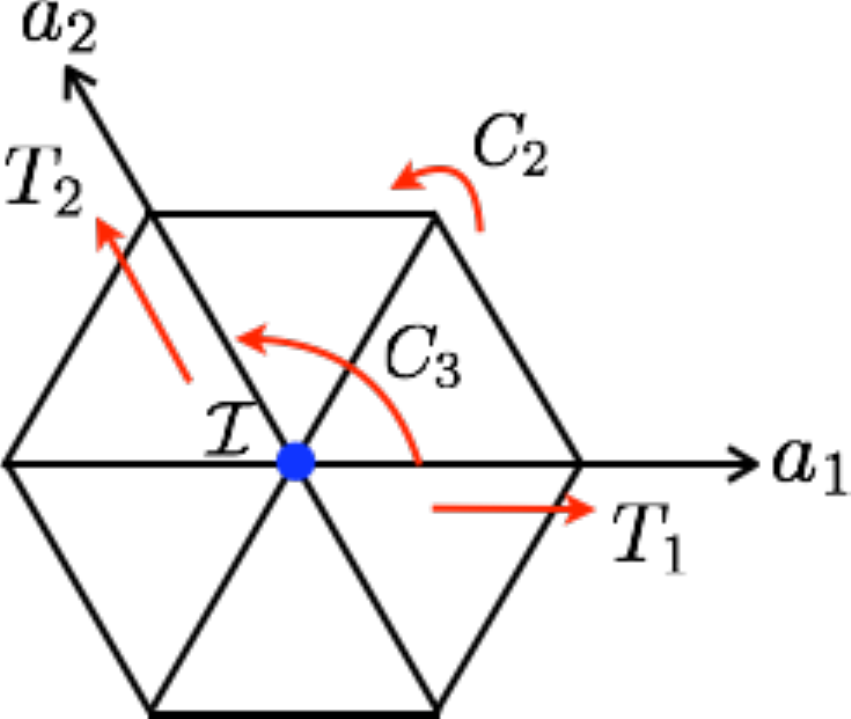} }
\caption{The space group symmetry operation for the Yb triangular layer. }
\label{fig3}
\end{figure}

In YbMgGaO${}_4$, the Yb$^{3+}$ ions form a perfect triangular lattice. 
The interlayer separation between nearby Yb triangular layers 
is 8.4$\AA$ and is much larger than the intralayer Yb lattice constant that 
is 3.4$\AA$~\cite{Yuesheng2015}. 
Because the Yb $4f$ electron is very localized spatially,
one can safely neglect the interlayer coupling and 
focus on the intralayer coupling between the Yb local moments. 
We thus keep the symmetry operation of the space group within each triangular layer.
As we show in Fig.~\ref{fig3}, the R$\bar{3}$m space group of YbMgGaO${}_4$ 
contains two translations, $T_1$ and $T_2$, along the two crystallographic
axes, the three-fold rotation, $C_3$, around the $z$ direction, 
the two-fold rotation, $C_2$ around the diagonal direction,
and an inversion, $\mathcal{I}$, about the triangular lattice site. 
With these symmetries and their transformations on the 
spin operators, it is ready to obtain the generic spin 
Hamiltonian in Eq.~(\ref{eq1}) that describes the interaction between 
the local moments. 

\section{Semiclassical analysis: Luttinger-Tisza method and classical Monte Carlo simulation}
\label{sec3}
 
To obtain the first understanding of the ground state properties 
of the generic spin model, in this section we will implement the 
standard Luttinger-Tisza method and classical Monte Carlo simulation 
to unconver the magnetic ordered ground states and to obtain the 
classical ground state phase diagram. 
 
\subsection{Luttinger-Tisza method}

Here we treat the effective spin ${\bf S}_i$ as a classical vector 
that satisfies the hard spin constraint $|{\bf S}_i | = 1/2$. 
Following Luttinger and Tisza~\cite{PhysRev.70.954}, 
we first replace the hard spin constraint with a global 
constraint such that 
\begin{equation}
\sum_i |{\bf S}_i |^2 = \frac{N}{4} ,
\end{equation}
where $N$ is the total number of spins. The classical spin Hamiltonian 
is then minimized under this global constraint. 
If the energy minimum turns out to satisfy the local 
hard spin constraint as well, then this energy minimum is 
the true classical ground state. 

There are four parameters, $J_{zz}, J_{\pm}, J_{\pm\pm}, J_{z\pm}$, 
in the generic spin model. We first consider the parameter regime 
when the anisotropic interaction vanishes with 
$J_{\pm\pm} = 0$ and $J_{z\pm} = 0$. In this regime the spin model 
reduces to the XXZ model. From the Curie-Weiss temperature results on 
single crystal YbMgGaO${}_4$ samples~\cite{Yuesheng2015}, 
one finds that both $J_{zz}$ and $J_{\pm}$ are antiferromagnetic 
and $J_{\pm}/J_{zz}\approx 0.915$
which is fixed to this value throughout the paper.
The ground state of this XXZ model is simply the well-known
120$^\circ$ ordered state with the spins orienting in the $xy$ plane.
The ordering wavevector of the 120$^\circ$ 
state is at 
\begin{equation}
{\bf k}_{\text c} = (\frac{4\pi}{3},0), 
\end{equation}
or its symmetry equivalent wavevectors. 

Now we discuss the effect of the anisotropic spin interactions. 
With a small $|J_{\pm\pm}|$, the minimum of the classical 
Hamiltonian under the global constraint slightly deviates 
from the 120$^\circ$ state and occurs at incommensurate wavevectors. 
In strong spin-orbit coupled insulators, however, the 
incommensurate ordering is generically not favored. 
Because of the intrinsic spin anisotropy that originates 
from the strong spin-orbit coupling~\cite{PhysRevB.92.020411}, 
to optimize the spin anisotropy, 
the ordered spin moments cannot orient freely like the case 
for an incommensurate state. As a result, we generically have 
the commensurate spin orders in the strong spin-orbit coupled insulators.
Apart from the general understanding, we here provide more specific reasons. 
Due to the low symmetry of the spin Hamiltonian, the eigenstate that corresponds
to the minimum is generically unique, hence one cannot find two orthogonal 
eigenvectors to construct an incommensurate spiral state that satisfies 
the hard spin constraint on every lattice site. 
Therefore, the incommensurate state cannot be a true classical ground state, 
and we tentatively regard the 120$^\circ$ state
as the candidate classical ground state in the 
regime with a small $J_{\pm\pm}$. 

\begin{figure}[t]
{\includegraphics[width=8cm]{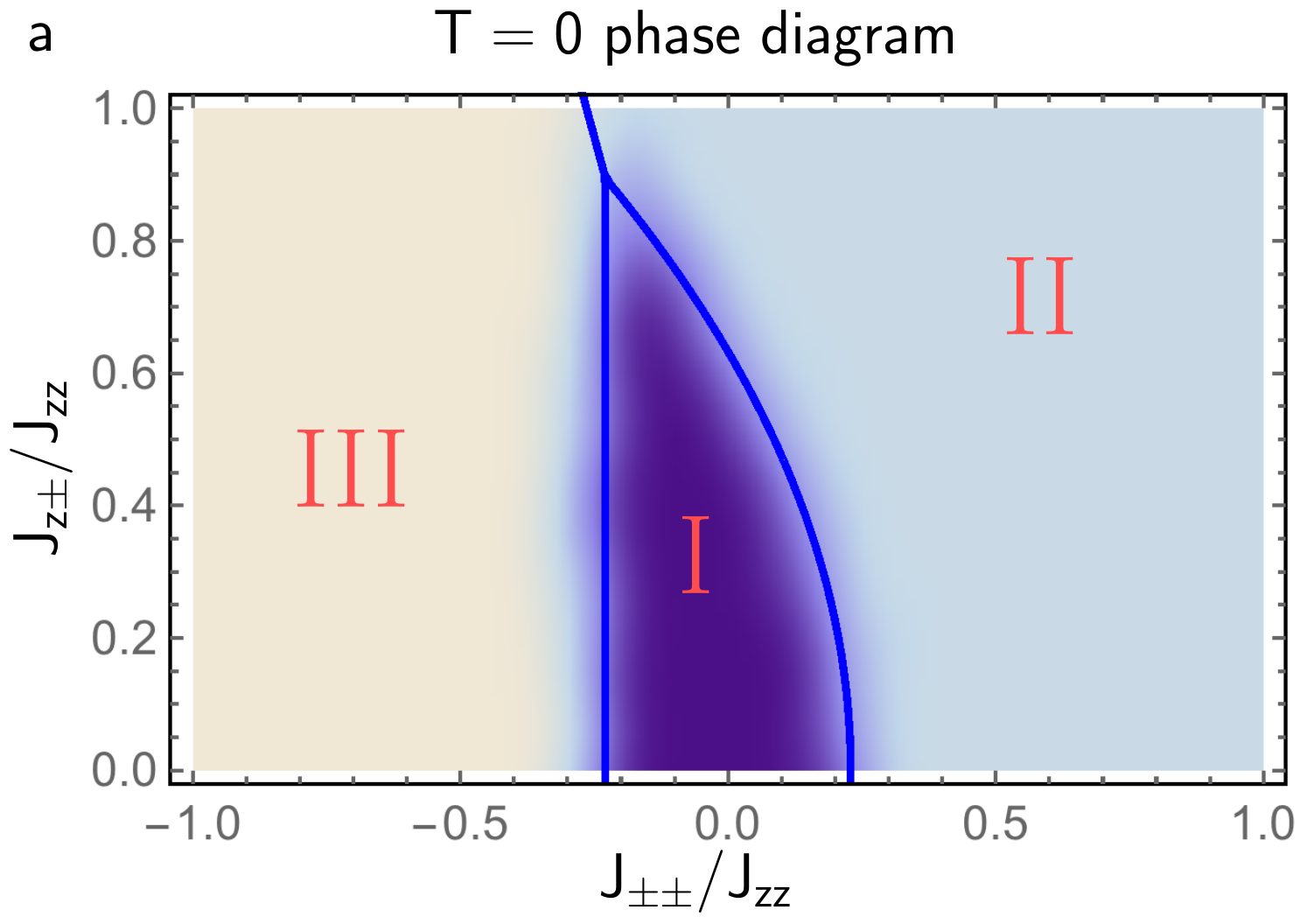}}
{\includegraphics[width=4cm]{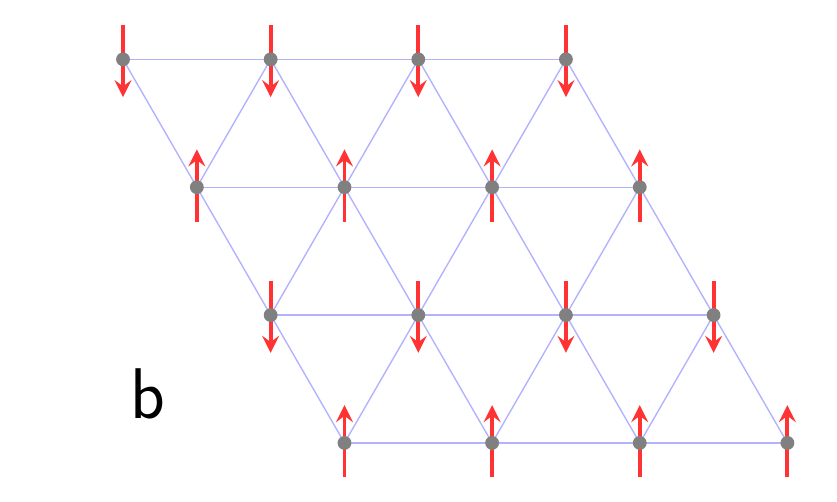}}
{\includegraphics[width=4cm]{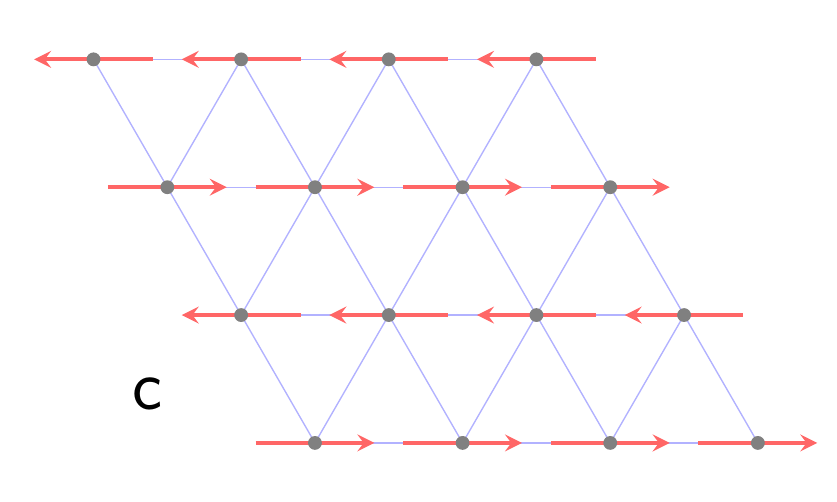}}
\caption{(Color online.) (a) The classical phase diagram in the zero temperature limit.
The solid phase boundaries determined by the Luttinger-Tisza method,
and the colored regions are determined by classical Monte Carlo simulation.
    (b) The stripe order in regin II with spins pointing in the $yz$ plane.
    (c) The stripe order in region III with spins pointing along the $x$ direction.}
\label{fig4}
\end{figure}

With a large $|J_{\pm\pm}|$ and/or a large $|J_{z\pm}|$, 
the minimum of the classical spin Hamiltonian occurs at 
\begin{equation}
{\bf k}_{\text s} = (0, \frac{2\pi}{\sqrt{3}} ), 
\end{equation}
or its symmetry equivalent wavevectors. Remarkably, 
this minimum state satisfies the hard spin 
constraint and is thus a true ground state. 
The spin configuration with this ordering 
wavevector has a stripe order, {\it i.e.}, the spins order ferromagnetically 
along one lattice direction and antiferromagnetically
along the remaining two lattice directions. 
To obtain the classical phase diagram in Fig.~\ref{fig4}a, 
we compare the energies of the 120$^\circ$ state and the stripe ordered phases.
In the region I of the phase diagram, the 120$^\circ$ state is obtained.
In the region II and III, we find two stripe ordered phases with different spin
orientations. Without loss of generality, we fix the ordering wavevector 
of the stripe phase to be ${\bf k}_{\text s}=(0, 2\pi/\sqrt{3})$. 
Due to the locking of the spin orientation and the
ordering wavevector, the spin configuration is fixed as well. 
With this choice of the ordering wavevector, the spins are pointing 
in the $yz$ plane~\footnote{The actual spin orientation depends on the 
couplings.} and $x$ direction in region II and region III, respectively
(see Fig.~\ref{fig4}). 

Here we elucidate the structure of the classical ground state phase diagram. 
The magnetic phases for a negative $J_{z\pm}$ 
can be simply generated from the ones in the positive $J_{z\pm}$ case 
by a 180$^\circ$ rotation around the $z$ axis in the spin space. 
Under this spin rotation, 
\begin{eqnarray}
S^z_i &\rightarrow &  S^z_i, \\
S^{\pm}_i & \rightarrow & -S^{\pm}_i,
\end{eqnarray}
the coupling $J_{z\pm} \rightarrow -J_{z\pm}$ 
while other couplings stay invariant~\cite{Savary12}. 
Therefore, we only consider the phase diagram with 
a positive $J_{z\pm}$ in Fig.~\ref{fig4}a. 
In addition, on the horizontal axis with $J_{z\pm} = 0$, 
the magnetic phases are symmetric about the origin. 
This is seen by rotating the spins around the $z$ axis by 90$^\circ$.
It transforms the spins as 
\begin{eqnarray}
S^z_i &\rightarrow & S^z_i \\
S^{\pm}_i &\rightarrow & \pm i S^{\pm}_i,
\end{eqnarray}
and the coupling as $J_{\pm\pm} \rightarrow -J_{\pm\pm}$.
The above properties of the classical phase diagram 
hold even for the quantum case. 

\subsection{Classical Monte Carlo simulation}
\label{sec3b}

To further investigate the structure of the classical phase diagram 
and to extract finite-temperature magnetic properties, 
we implement the classical Monte Carlo simulation of 
the classical spin Hamiltonian~\cite{MonteCarlo1, MonteCarlo2}. 
As we previously explained, the system prefers the commensurate
spin orders. So one does not need a large system size 
to carry out the classical Monte Carlo simulation. 
The simulation is performed on $6\times 6$ and $12 \times 12$ 
triangular systems. It starts with a randomly chosen initial 
spin configuration, followed by $5000$ transient Monte Carlo steps (MCS) for 
the system to equilibrate. Within each step, the Metropolis 
algorithm~\cite{Metropolis1, Metropolis2} is implemented for sampling, 
and a method proposed in Ref.~\onlinecite{SpherePoint} 
is used for updating the spin configurations in the canonical ensemble. 
The observables are averaged within a sample of size MCS $=50000$.

\begin{figure*}[t]
{\includegraphics[width=.3\textwidth]{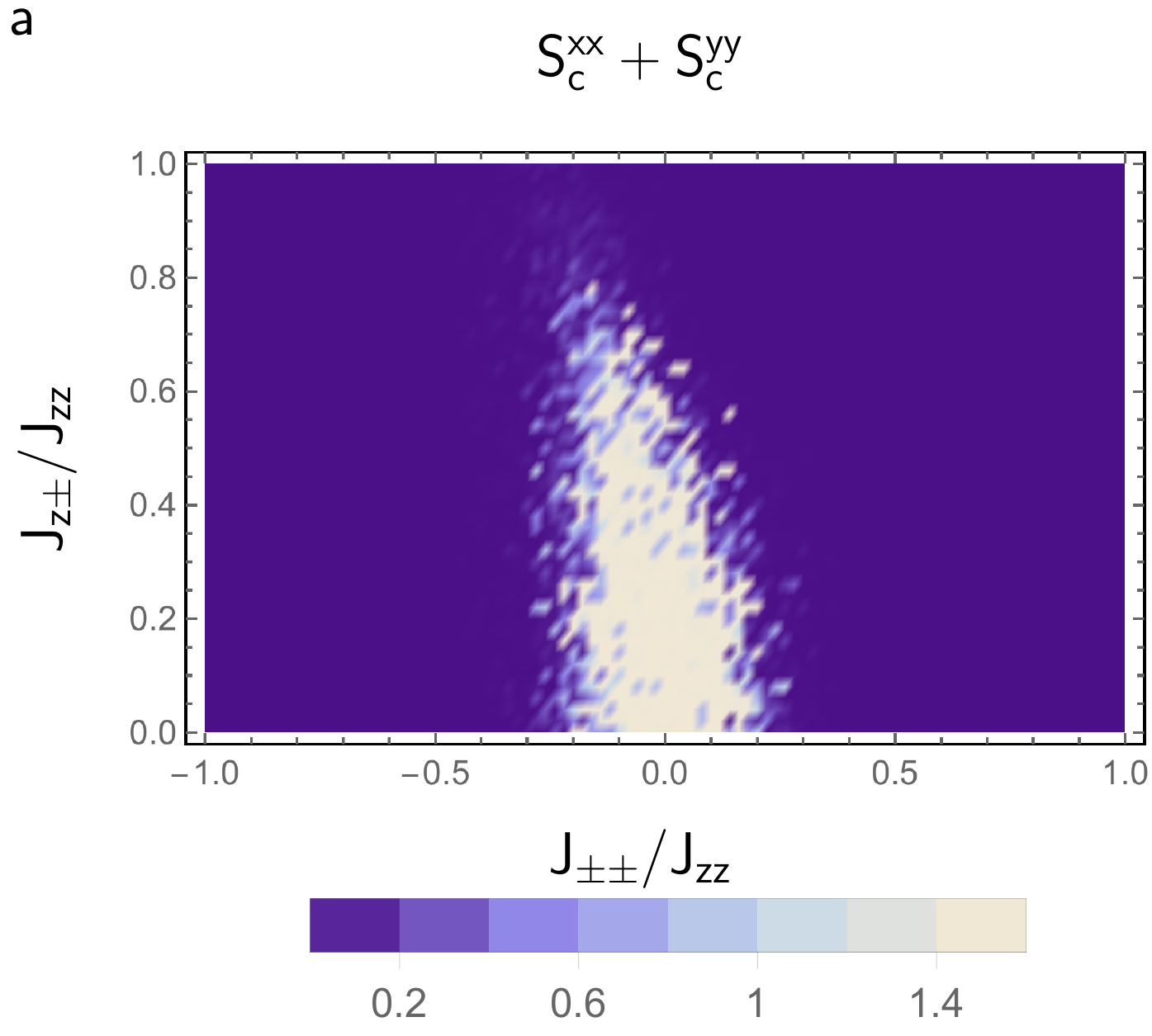}}
{\includegraphics[width=.3\textwidth]{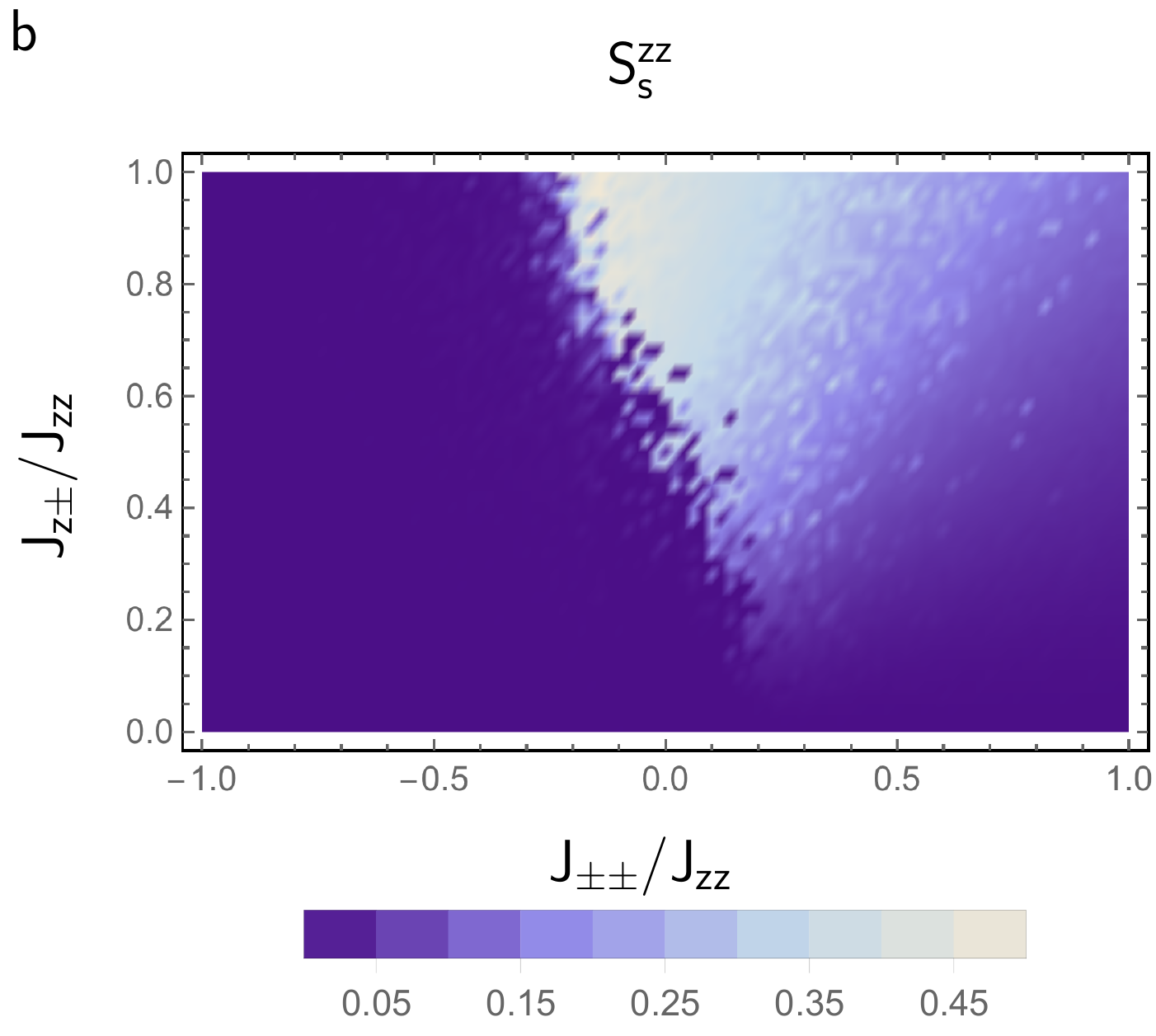}}
{\includegraphics[width=.3\textwidth]{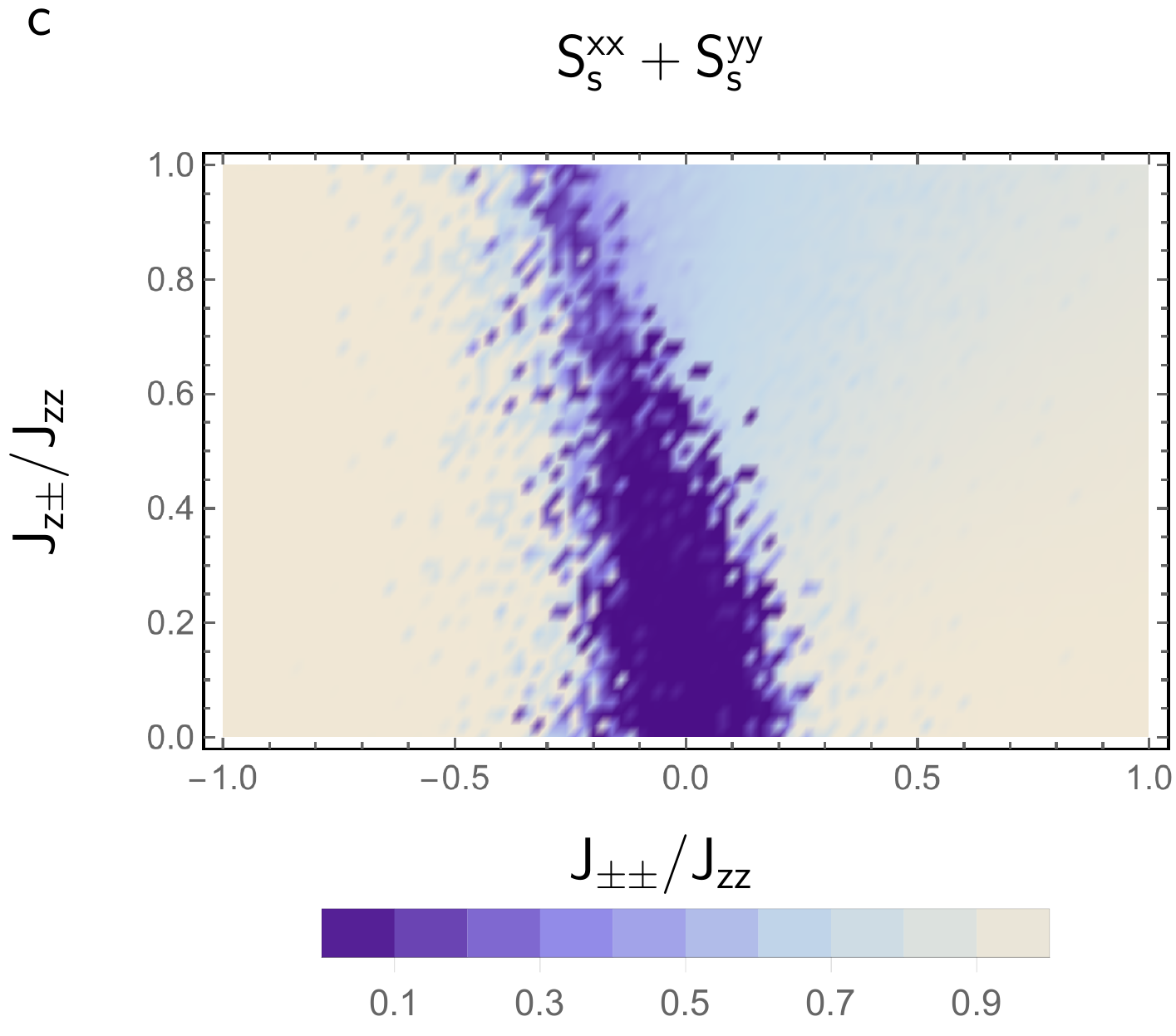}}
\caption{(Color online.) 
 Spin-spin correlation functions $\mathsf{S}^{\alpha \beta}_{c/s}$ at zero temperature are displayed.
(a) $\mathsf{S}^{x x}_{c} + \mathsf{S}^{y y}_{c}$ at $T\rightarrow 0$.
 The region with large $\mathsf{S}_c$ suggests the 120$^\circ$ order of spins.
(b) A finite $z$-$z$ correlation for the stripe phase helps us distinguish two 
stripe phases (Fig.~\ref{fig4}) due to different signs of $J_{\pm\pm}$.
(c) $x$-$x$ plus $y$-$y$ correlation for the stripe phase.
\label{fig5}
}
\end{figure*}

Since the 120$^\circ$ state (the stripe ordered phase) has   
an ordering wavevector ${\bf k}_{\text c} $ (${\bf k}_{\text s}$), 
we evaluate the spin-spin correlation functions at the corresponding
wavevectors,
\begin{equation}
	\mathsf{S}^{\alpha \beta}_{c/s} = \frac{1}{N^2} \sum_{i,j} 
	\langle{S_i^\alpha S_j^\beta} \rangle
	e^{i{\bf k}_{\text{c/s}} \cdot \lb {\bf r}_j - {\bf r}_i \rb},
\end{equation}
where $\alpha, \beta = x, y, z$. The result is summarized in Fig.~\ref{fig5}. 
In the zero temperature limit, we observe a significant stripe order 
that signifies the stripe phases away from the central region of the phase diagram. 
We also notice that as both $J_{\pm\pm}$ and $J_{z\pm}$   
increase on the positive side, the spins develop a finite 
component in the $yz$ plane, distinguishing it from the 
stripe order with spins pointing along $x$ direction 
in the negative $J_{\pm\pm}$ region (see Fig.~\ref{fig4}).  

Near the phase boundaries, not only the neighboring ordered 
phases are very close in energies, 
but a large number of classical spin configurations have
rather close energies. As a result, thermal fluctuations   
can easily populate the low energy spin configurations
even at a temperature much smaller than $|\Theta_{\text{CW}}|$ 
such that the system may not favor any obvious magnetic order. 
Therefore, we expect the ordering temperature 
to be strongly suppressed in these frustrated regions.  

\begin{figure*}[t]
 {\includegraphics[height=.3\textwidth]{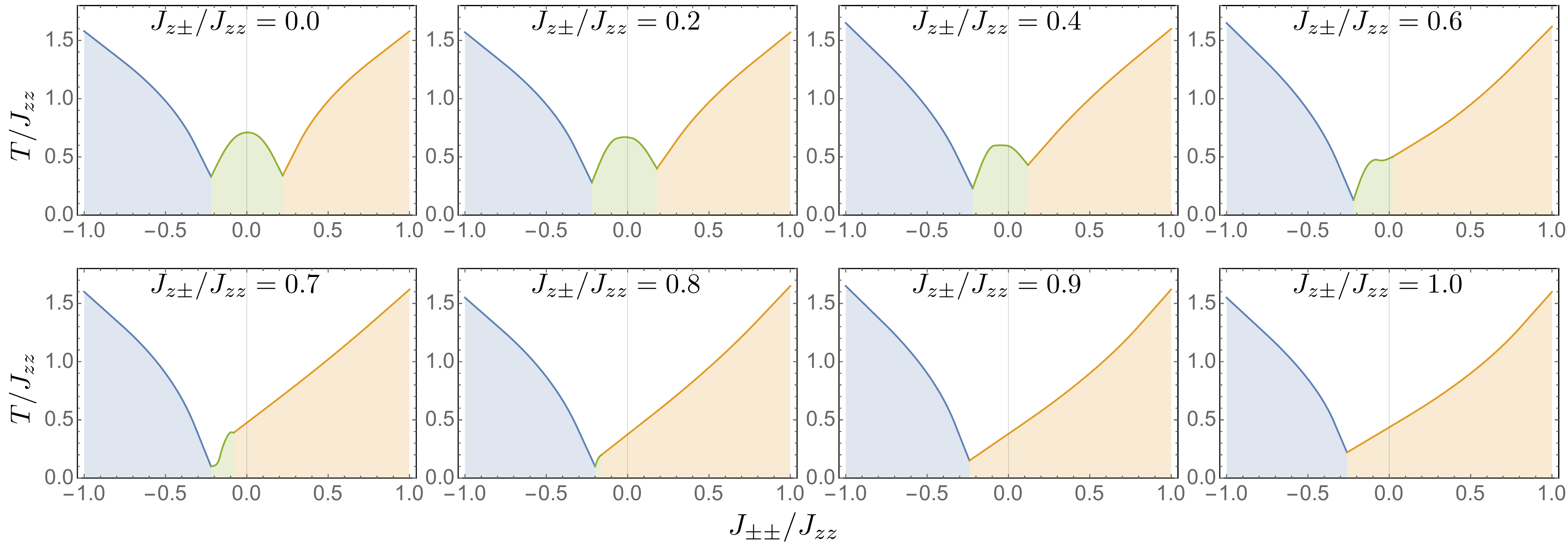}}
\caption{
Cuts through constant $J_{z\pm}$ lines of through the three-dimensional
finite temperature phase diagram in the main text. The white region 
is the high temperature paramagnetic phase. The solid line indicates
the transition temperatures. 
 }
\label{fig6}
\end{figure*}

The classical Monte Carlo simulation allows us to access the finite 
temperature magnetic properties. We can still perform the calculation 
of the spin correlation function as the temperature is raised from zero. 
At zero temperature, the system is frozen at its ground state, therefore
the deviation of a physical observable $\hat{O}$ 
(chosen to be $\mathsf{S}^{\alpha \beta}$ in this case),
$\langle{\hat{O}^2}\rangle - \langle \hat{O} \rangle ^2 $, vanishes. 
However, at finite temperatures, due to the possibility for 
spins to flip to another configuration with
similar energy, $\hat{O}$ can develop a nonzero deviation.
Therefore, the Binder ratios~\cite{BinderRatio1,BinderRatio2}, 
defined for the spin-spin correlation functions,
\be
r^{\alpha\alpha} = 
\frac{ \langle  ( \mathsf{S}_{c/s}^{\alpha\alpha} ) ^2 \rangle}
{ \langle \mathsf{S}_{c/s}^{\alpha\alpha} 
\rangle ^2},
\ee
should attain the value $1$ at zero temperature, 
and saturate to a larger value in the high temperature limit. 
The Binder ratios are scale-independent quantities 
at the critical temperature $T_c$, hence $T_c$ can be 
estimated by finding the crossing of $r^{\alpha\alpha}$-$T$ 
curves for different lattice sizes. The thermal transition is found to be continuous and 
no other thermal phases are found in our numerical 
study of finite size systems. The result of our simulation 
is summarized in Fig.~\ref{fig6}. 

\begin{figure*}[t]
 {\includegraphics[height=.3\textwidth]{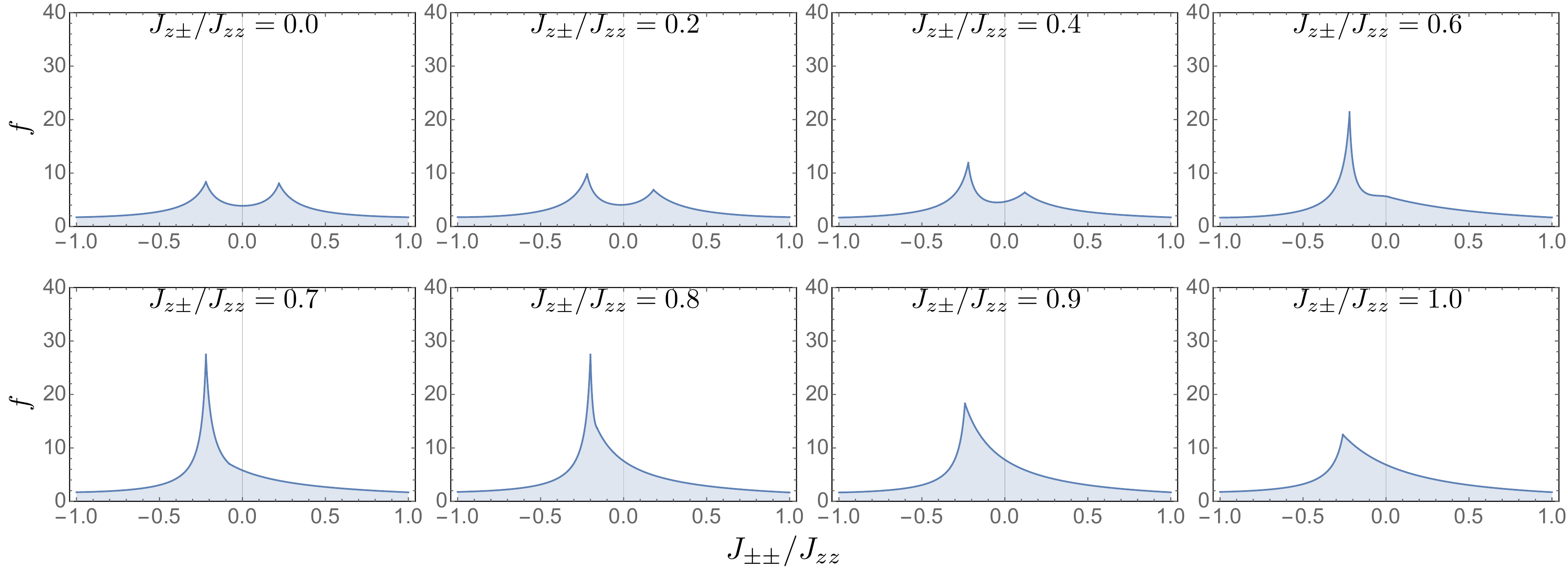}}
\caption{(Color online.) The frustration parameter $f\equiv |\Theta_{\text{CW}}|/T_{\text N}$ of 
the corresponding plots in Fig.~\ref{fig6}.}
\label{fig7}
\end{figure*}

It is found that in the parameter regimes near the phase boundary
the magnetic ordering temperature is in fact strongly suppressed 
compared to the Curie-Weiss temperature $|\Theta_{\text{CW}}|$. 
The local moments do not order down to very low temperatures, which
indicates the strong spin frustration in these regions. 
In Fig.~\ref{fig7}, we evaluate the frustration parameter 
\begin{equation}
f\equiv \frac{|\Theta_{\text{CW}}|}{T_c}
\end{equation} 
that is an empirical measure of the frustration~\cite{Ramirez1994}. 
Because of the spin anisotropy, the Curie-Weiss temperature depends
on the direction of the external magnetic field. 
To be specific, we choose the Curie-Weiss temperature to be the one
when the external field is applied in the $xy$ plane, so 
we have $\Theta_{\text{CW}} = -3J_{\pm}$~\cite{Yuesheng2015}. 
Indeed, the frustration parameter is as large as 20 
near the phase boundaries between two neighboring
phases (see Fig.~\ref{fig7}).

\section{Quantum fluctuation and self-consistent Dyson-Maleev spin wave theory}
\label{sec4}

The semiclassical analysis in the previous section 
gives the classical ground state phase diagram. 
When the ground state does support magnetic ordering, 
the semiclassical treatment does provide a qualitative 
understanding of the magnetic phases. In this section, 
we go beyond the semiclassics and 
access the quantum mechanical nature of the local moments. 
By considering the quantum fluctuation in the magnetic ordered phases, 
we try to understand the stability of  
the magnetic order in the presence of quantum fluctuations and 
find out where the disordered state may occur in the phase diagram. 

Deep inside each ordered phase, the magnetic order is clearly very robust, 
and we expect that the quantum fluctuation would merely renormalize 
the magnetic order. In contrast, near the phase boundary,
many classical spin configurations have rather close energies and 
may strongly enhance the quantum fluctuations. 
To demonstrate this explicitly, we apply the Dyson-Maleev transformation 
for the spin operators and solve for the quantum correction
to the magnetic order within a self-consistent 
spin wave theory~\cite{PhysRev.102.1217,Maleev}. To be specific,
we focus the analysis on the stripe ordered phase in region III, 
and the spin wave theory in other ordered regions can be obtained likewise. 
As we show in Sec.~\ref{sec3}, the spins in region III orient in the 
$\pm \hat{x}$ directions. We introduce the Dyson-Maleev representation 
for the spin operators~\cite{PhysRev.102.1217,Maleev}
\begin{eqnarray}
	{\bf S}_i \cdot \hat{m}_i &=& S - a^\dagger_i a^{\phantom\dagger}_i, \label{eq3}\\
	{\bf S}_i \cdot [\hat{m}_i \times \hat{z}] &=& \frac{1}{2}
	[ a^\dagger_i (2S - a^\dagger_i a^{\phantom\dagger}_i)
	+ a^{\phantom\dagger}_i ],\label{eq4} \\
	{\bf S}_i \cdot \hat{z} &=& 
\frac{1}{2i} 	[a^\dagger_i (2S - a^\dagger_i a^{\phantom\dagger}_i)
  - a^{\phantom\dagger}_i]	,\label{eq5}
\end{eqnarray}
where the spin magnitude $S=1/2$, 
$\hat{m}_i$ is the direction of the classical spin order 
and orients along $\hat{x}$ or $-\hat{x}$. 
Because the stripe ordered state  
has two magnetic sublattices, there are two flavors of Dyson-Maleev bosons that 
describe the magnetic excitation and quantum fluctuation in region III. 

In the usual linear spin wave approximation, one neglects the cubic 
boson terms in the Dyson-Maleev transformation by setting 
\begin{eqnarray}
{\bf S}_i \cdot \hat{m}_i &=& S - a^\dagger_i a^{\phantom\dagger}_i,  \\
{\bf S}_i \cdot [\hat{m}_i \times \hat{z}] &\approx& (a^\dagger_i + a^{\phantom\dagger}_i )/2 
\\
{\bf S}_i \cdot \hat{z} &\approx &
 (a^\dagger_i - a^{\phantom\dagger}_i) /(2i) .
\end{eqnarray}
This approximation is valid when $\langle a^\dagger_i a^{\phantom\dagger}_i 
\rangle \ll S$. We substitute the spin operators with the Dyson-Maleev bosons, 
keep the quadratic part of the spin-wave Hamiltonian, 
and diagonalize it with the standard Bogoliubov transformation. 
We proceed to evaluate the quantum correction 
$\delta m_i\equiv \langle a_i^\dagger a_i^{\phantom\dagger} \rangle$ 
and find that $\delta m_i$ is comparable to the spin magnitude 
in the parameter regime near the phase boundary. Clearly, the strong 
quantum fluctuation in these regions invalidates the assumption 
of the linear spin wave theory that neglects the boson interaction
in the formalism.

To fix the drawbacks of the linear spin wave approximation, 
we implement a self-consistent spin wave calculation in the following. 
The Dyson-Maleev transformation in Eqs.~(\ref{eq3})-(\ref{eq5}) has proven to be 
convenient for studying spin-wave interaction~\cite{PhysRevB.45.10131}. 
With the Dyson-Maleev transformation for the spin operators, we 
obtain the spin wave Hamiltonian. In this Hamiltonian, there exist cubic, quartic, quintic,  
and sextic terms in terms of the Dyson-Maleev bosons. 
To reduce the spin wave Hamiltonian down 
to the quadratic level, we make the mean-field decoupling of 
the quartic and sextic terms. The quartic and 
sextic terms are decoupled into 
various on-site and intersite boson bilinears, 
\begin{widetext}
\begin{eqnarray}
a^\dagger_i a^{\phantom\dagger}_i a^\dagger_j a^{\phantom\dagger}_j
& \rightarrow &
[a^\dagger_i a^{\phantom\dagger}_i - \langle a^\dagger_i a^{\phantom\dagger}_i \rangle]  
\langle a^\dagger_j a^{\phantom\dagger}_j\rangle
+ \langle a^\dagger_i a^{\phantom\dagger}_i \rangle 
[a^\dagger_j a^{\phantom\dagger}_j - \langle a^\dagger_j a^{\phantom\dagger}_j  \rangle ] 
\nonumber \\
&+& \langle a^\dagger_i a_j^\dagger \rangle  [a^{\phantom\dagger}_i a^{\phantom\dagger}_j - 
\langle a^{\phantom\dagger}_i a^{\phantom\dagger}_j \rangle ]
+ [a^\dagger_i a_j^\dagger - \langle a^\dagger_i a_j^\dagger\rangle ]
 \langle a^{\phantom\dagger}_i a^{\phantom\dagger}_j\rangle
\nonumber \\ 
& + & \langle a^\dagger_i a^{\phantom\dagger}_j \rangle 
[ a^\dagger_j a^{\phantom\dagger}_i - \langle a^\dagger_j a^{\phantom\dagger}_i \rangle]
+  [ a^\dagger_i a^{\phantom\dagger}_j -\langle a^\dagger_i a^{\phantom\dagger}_j\rangle]
 \langle a^\dagger_j a^{\phantom\dagger}_i \rangle
\nonumber \\
&+& \langle a^\dagger_i a^{\phantom\dagger}_i \rangle \langle a^\dagger_j a^{\phantom\dagger}_j \rangle
+ \langle a^\dagger_i a_j^\dagger \rangle  \langle a^{\phantom\dagger}_i a^{\phantom\dagger}_j \rangle
+ \langle a^\dagger_i a^{\phantom\dagger}_j \rangle \langle a^\dagger_j a^{\phantom\dagger}_i \rangle ,
\\
a^{\dagger}_i a_j^{\dagger} a_j^{\dagger} a_j^{\phantom\dagger}
& \rightarrow & 
[a_j^{\dagger}a_j^{\dagger}-\langle a_j^{\dagger}a_j^{\dagger}\rangle]
\langle a^{\dagger}_i a_j^{\phantom\dagger} \rangle 
+ a_j^{\dagger}a_j^{\dagger} 
[ a^{\dagger}_i a_j^{\phantom\dagger} - \langle a^{\dagger}_i a_j^{\phantom\dagger} \rangle]
\nonumber \\
&+&
2[a_j^{\dagger}a_j^{\phantom\dagger} 
- \langle a_j^{\dagger}a_j^{\phantom\dagger} \rangle ]
\langle a_i^{\dagger} a_j^{\dagger} \rangle 
+ 2 \langle a_j^{\dagger}a_j^{\phantom\dagger} \rangle 
[a_i^{\dagger} a_j^{\dagger} - \langle a_i^{\dagger} a_j^{\dagger} \rangle ]
\nonumber \\
&+& 
\langle a_j^{\dagger}a_j^{\dagger} \rangle
\langle a^{\dagger}_i a_j^{\phantom\dagger} \rangle
+ 2 \langle a_j^{\dagger}a_j^{\phantom\dagger} \rangle 
\langle a_i^{\dagger} a_j^{\dagger} \rangle,
\\
a^\dagger_i a^\dagger_i a^{\phantom\dagger}_i a^\dagger_j a^\dagger_j a^{\phantom\dagger}_j
& \rightarrow &
[ a^\dagger_i a^\dagger_i - \langle a^\dagger_i a^\dagger_i \rangle ]
\langle a_i^{\phantom\dagger} a_j^\dagger a_j^\dagger a_j^{\phantom\dagger} \rangle
+ 2 [ a^\dagger_i a^{\phantom\dagger}_i - \langle a^\dagger_i a^{\phantom\dagger}_i \rangle ] 
\langle a_i^{\dagger} a_j^\dagger a_j^\dagger a_j^{\phantom\dagger} \rangle
\nonumber \\
&+&
[ a^\dagger_j a^\dagger_j - \langle a^\dagger_j a^\dagger_j  \rangle ]
\langle a_i^{\dagger} a_i^\dagger a_i^{\phantom\dagger} a_j^{\phantom\dagger} \rangle
+ 2 [ a^\dagger_j a^{\phantom\dagger}_j -\langle a^\dagger_j a^{\phantom\dagger}_j \rangle ]
\langle a_i^{\dagger} a_i^\dagger a_i^{\phantom\dagger} a_j^{\dagger} \rangle
\nonumber \\
&+&
4 [a^\dagger_i a^{\dagger}_j - \langle a^\dagger_i a^{\dagger}_j \rangle]
  \langle a_i^\dagger a_i^{\phantom\dagger} a_j^\dagger a_j^{\phantom\dagger} \rangle
+ 2 [a^\dagger_i a^{\phantom\dagger}_j - \langle a^\dagger_i a^{\phantom\dagger}_j  \rangle ]
  \langle a_i^\dagger a_i^{\phantom\dagger} a_j^\dagger a_j^{\dagger}  \rangle
  \nonumber \\
&+& 2 [a^{\phantom\dagger}_i a^{\dagger}_j -
\langle  a^{\phantom\dagger}_i a^{\dagger}_j  \rangle ]
  \langle a^\dagger_i a^\dagger_i a^\dagger_j a^{\phantom\dagger}_j    \rangle
+ [ a^{\phantom\dagger}_i a^{\phantom\dagger}_j 
- \langle a^{\phantom\dagger}_i a^{\phantom\dagger}_j \rangle ]
  \langle a^\dagger_i a^\dagger_i a^\dagger_j a^{\dagger}_j   \rangle   
\nonumber \\
&+& \langle a^\dagger_i a^\dagger_i a^{\phantom\dagger}_i a^\dagger_j a^\dagger_j a^{\phantom\dagger}_j
\rangle, 
\end{eqnarray}
where the expectation ``$\langle \cdots \rangle$'' 
is evaluated with respect to the ground 
state of the quadratic spin wave Hamiltonian that is defined later, and
we have 
\begin{eqnarray}
\langle a^\dagger_i a^\dagger_i a^{\phantom\dagger}_i a^\dagger_j a^\dagger_j a^{\phantom\dagger}_j
\rangle
&=&
2 \langle a^\dagger_i a^\dagger_i \rangle 
  \langle a^{\phantom\dagger}_i a^\dagger_j \rangle
  \langle a^{\dagger}_j a^{\phantom\dagger}_j \rangle
  + 
  \langle a^\dagger_i a^\dagger_i \rangle 
  \langle a^{\dagger}_j a^\dagger_j \rangle
  \langle a^{\phantom\dagger}_i a^{\phantom\dagger}_j \rangle
  + 4  \langle a^\dagger_i a^{\phantom\dagger}_i \rangle 
  \langle a^{\dagger}_j a^{\phantom\dagger}_j \rangle
  \langle a^{\dagger}_i a^{\dagger}_j \rangle
  \nonumber \\
  &+& 
  2 \langle a^\dagger_i a^{\phantom\dagger}_i \rangle 
  \langle a^{\dagger}_j a^{\dagger}_j \rangle
  \langle a^{\dagger}_i a^{\phantom\dagger}_j \rangle
  + 4 \langle a^\dagger_i a^{\dagger}_j \rangle 
  \langle a^{\dagger}_i a^{\phantom\dagger}_j \rangle
  \langle a^{\phantom\dagger}_i a^{\dagger}_j \rangle
  + 2 \langle a^\dagger_i a^{\dagger}_j \rangle 
  \langle a^{\dagger}_i a^{\dagger}_j \rangle
  \langle a^{\phantom\dagger}_i a^{\phantom\dagger}_j \rangle ,
\end{eqnarray}
\end{widetext}
and 
\begin{eqnarray}
\langle 
a_i^{\phantom\dagger} a_j^\dagger a_j^\dagger a_j^{\phantom\dagger} \rangle
&=&
2\langle a_i^{\phantom\dagger} a_j^\dagger \rangle 
\langle a_j^\dagger a_j^{\phantom\dagger} \rangle
+ 
\langle a_i^{\phantom\dagger} a_j^{\phantom\dagger}  \rangle 
\langle a_j^\dagger  a_j^\dagger \rangle ,
\\
\langle a_i^{\dagger} a_j^\dagger a_j^\dagger a_j^{\phantom\dagger} \rangle 
&=&
2\langle a_i^{\dagger} a_j^\dagger \rangle \langle a_j^\dagger a_j^{\phantom\dagger}  \rangle
+
\langle a_i^{\dagger} a_j^{\phantom\dagger} \rangle 
\langle a_j^\dagger a_j^\dagger \rangle ,
\\
\langle a_i^{\dagger} a_i^\dagger a_i^{\phantom\dagger} a_j^{\phantom\dagger} \rangle
&=&
2 \langle a_i^\dagger a_i^{\phantom\dagger} \rangle 
 \langle  a_i^\dagger a_j^{\phantom\dagger} \rangle 
 + 
 \langle a_i^{\dagger} a_i^\dagger \rangle
 \langle a_i^{\phantom\dagger} a_j^{\phantom\dagger} \rangle ,
\\
\langle a_i^{\dagger} a_i^\dagger a_i^{\phantom\dagger} a_j^{\dagger} \rangle
&=& 
2 \langle a_i^\dagger a_i^{\phantom\dagger} \rangle \langle a_i^{\dagger} a_j^{\dagger} \rangle
+ \langle a_i^{\dagger} a_i^\dagger \rangle \langle a_i^{\phantom\dagger} a_j^{\dagger} \rangle,
\\
\langle a_i^\dagger a_i^{\phantom\dagger} a_j^\dagger a_j^{\phantom\dagger} \rangle
&=& \langle a_i^\dagger a_i^{\phantom\dagger} \rangle 
    \langle a_j^\dagger a_j^{\phantom\dagger} \rangle 
 +  \langle a_i^\dagger a_j^\dagger \rangle \langle a_i^{\phantom\dagger} a_j^{\phantom\dagger}\rangle
 \nonumber \\
 & & +  \langle a_i^\dagger a_j^{\phantom\dagger} \rangle 
        \langle a_j^\dagger a_i^{\phantom\dagger} \rangle ,
\\
\langle a_i^\dagger a_i^{\phantom\dagger} a_j^\dagger a_j^{\dagger}  \rangle
&=& 
\langle a_i^\dagger a_i^{\phantom\dagger} \rangle \langle a_j^\dagger a_j^{\dagger} \rangle 
+ 2 \langle a_i^\dagger a_j^\dagger \rangle
 \langle a_i^{\phantom\dagger} a_j^\dagger \rangle ,
\\
\langle a^\dagger_i a^\dagger_i a^\dagger_j a^{\phantom\dagger}_j    \rangle
&=&
 2 \langle a^\dagger_i a^\dagger_j \rangle \langle a^\dagger_i a^{\phantom\dagger}_j \rangle
 +
 \langle a^\dagger_i a^\dagger_i \rangle \langle a^\dagger_j a^{\phantom\dagger}_j \rangle ,
\\
  \langle a^\dagger_i a^\dagger_i a^\dagger_j a^{\dagger}_j   \rangle   
&=& 
 2 \langle a^\dagger_i a^\dagger_j \rangle \langle a^\dagger_i a^\dagger_j  \rangle 
+ \langle a^\dagger_i a^\dagger_i \rangle \langle a^\dagger_j a^{\dagger}_j   \rangle .
\end{eqnarray}

The decoupling of the cubic and quintic terms leads to linear terms 
in the Dyson-Maleev bosons that should all cancel out by the stability 
requirement of the classical ground state. 
Therefore, the decoupling of the cubic and quintic terms does not 
introduce extra quadratic terms into the spin-wave Hamiltonian. 

After defining the Fourier transform of the Dyson-Maleev boson operators, 
the quadratic spin-wave Hamiltonian can be organized as
\begin{eqnarray}
H_{\text{sw}} = \sum_{{\bf k} \in \text{BZ}'} 
(A^{\dagger}_{\bf k}, A_{-{\bf k}}^{\phantom\dagger})
\left( 
\begin{array}{ll}
F_{\bf k} & G^{\dagger}_{\bf k} \\
G_{\bf k} & F_{-{\bf k}}
\end{array}
\right)
\left(
\begin{array}{l}
A_{\bf k}\\
A^{\dagger}_{-{\bf k}}
\end{array}
\right),
\end{eqnarray}
where $A_{\bf k} 
= (a_{1{\bf k}}, a_{2{\bf k}})$ is the vector 
of the Dyson-Maleev boson annihilation operator, the 
subindices ``1'' and ``2'' label
the two sublattices of the magnetic unit cell,
and BZ$'$ is the magnetic Brioullin zone of the stripe ordered 
phase. $F_{\bf k}$ and $G_{\bf k}$ are $2 \times 2$
matrices and depend on the mean field parameters that
were introduced as boson bilinears. 
The quadratic spin wave Hamiltonian is diagonalized by 
the standard Bogoliubov transformation 
$Q_{\bf k}$~\cite{PhysRevB.76.064418},
\begin{equation}
\left(
\begin{array}{l}
B_{\bf k} \\ B_{-{\bf k}}^{\dagger}
\end{array}
\right)
 = Q_{\bf k}
 \left(
\begin{array}{l}
A_{\bf k} \\ A_{-{\bf k}}^{\dagger}
\end{array}
\right),
\end{equation}
where $B_{\bf k} = (b_{1{\bf k}}, b_{2{\bf k}})$ refers to the set of Bogoliubov bosons, 
and $Q_{\bf k}$ is a $4\times 4$ matrix that defines the Bogoliubov transformation. 
From the ground state of the quadratic spin wave Hamiltonian,
we evaluate the mean-field boson bilinears 
($\langle a^\dagger_i a^{\phantom\dagger}_i\rangle$,
$\langle a^\dagger_i a^{\phantom\dagger}_j\rangle$,
$\langle a^\dagger_i a^{\dagger}_i\rangle$, and
$\langle a^\dagger_i a^{\dagger}_j\rangle$).
As the spin wave Hamiltonian depends on these boson bilinears,
so we solve for them self-consistently
by an iteration method.

\begin{figure}[t]
{
 \includegraphics[width=8cm]{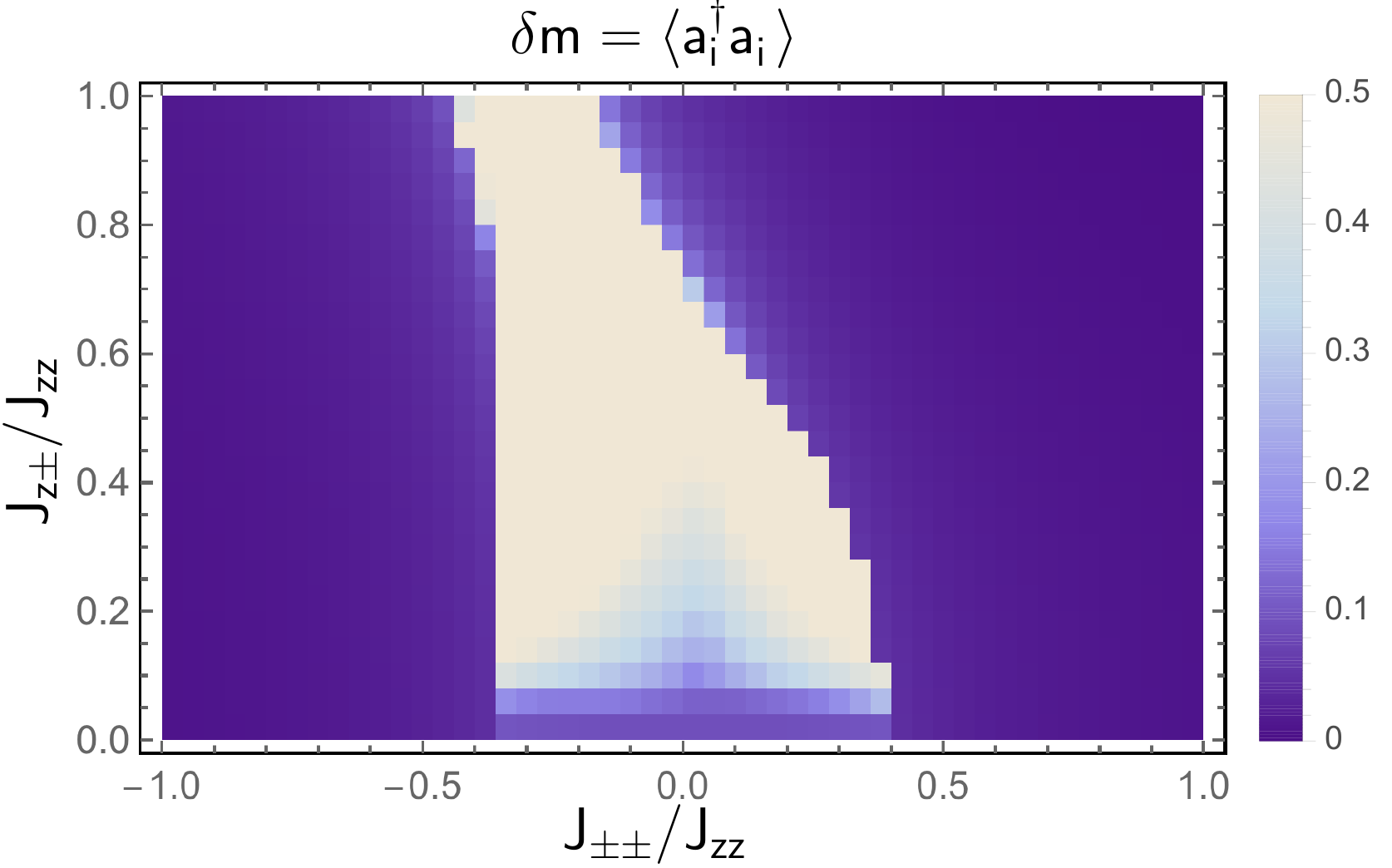}
}
\caption{(Color online.) Quantum correction ($\delta m$) to the magnetic orders 
that is calculated within the self-consistent spin wave theory 
on a $80\times80$ lattice. The region near phase boundary 
where $\delta m$ exceeds the spin magnitude with $\delta m \geq 1/2$ 
is marked in beige.
 } 
\label{fig8}
\end{figure}

The quantum correction to the magnetic order is evaluated by
\begin{eqnarray}
\delta m & = &  
 \langle a_i^\dagger a_i^{\phantom\dagger} \rangle =    \frac{1}{N} \sum_i 
\langle a_i^\dagger a_i^{\phantom\dagger} \rangle
\nonumber \\
& = & 
\frac{1}{2} \{ \frac{1}{N} \sum_{\bf k} \sum_{i=1}^2 
{[Q^\dagger_{\bf k} Q_{\bf k}^{\phantom\dagger}]}_{ii} - 1 \},
\label{scon}
\end{eqnarray}
where $N$ is the nubmer of lattice sites
and we have used the simple fact that the state in 
region III is invariant under the combined operation of 
time reversal and the translation $T_2$. 
If $\delta m > S$, the quantum fluctuation is very strong
and completely melts the magnetic order. 
As we show in Fig.~\ref{fig8}, the quantum fluctation is 
indeed quite strong and melts the magnetic order 
in the regions near the phase boundary. 
This suggests the ground state is likely to be disordered 
in these regions.

\section{Magnetic excitations with and without external magnetic fields} 
\label{sec5}

In this section, we study the properties of the magnetic excitations 
in different ordered phases as well as in the presence of strong magnetic fields.

\subsection{Linear spin wave theory for the three ordered phases}
\begin{figure}[t]
{
 \includegraphics[width=8cm]{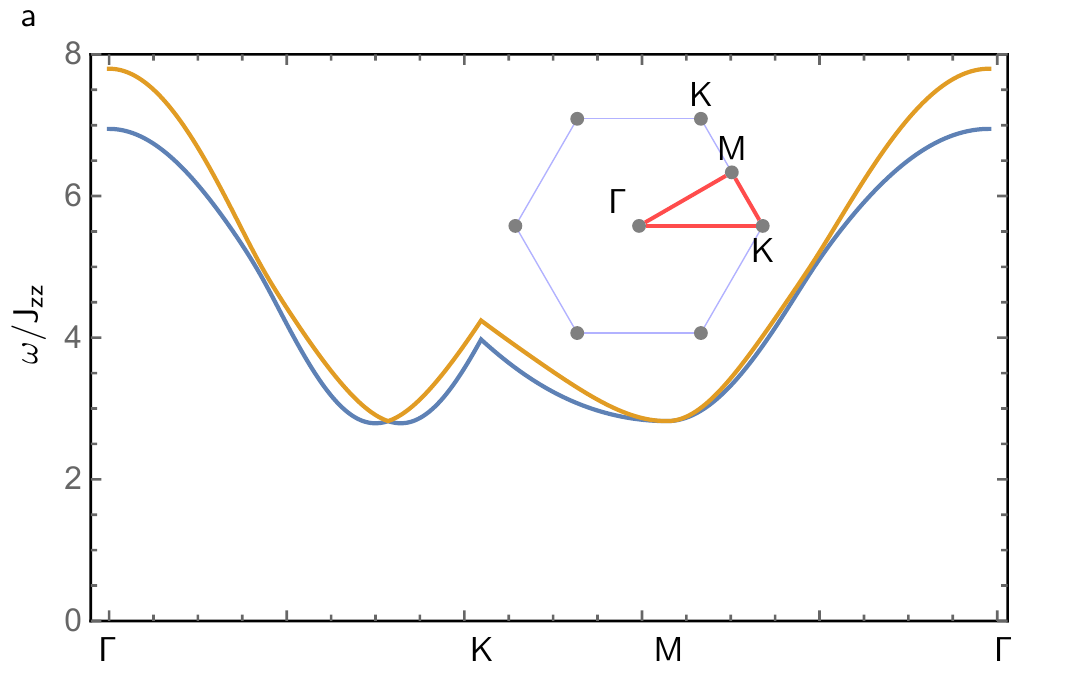}
 \includegraphics[width=8cm]{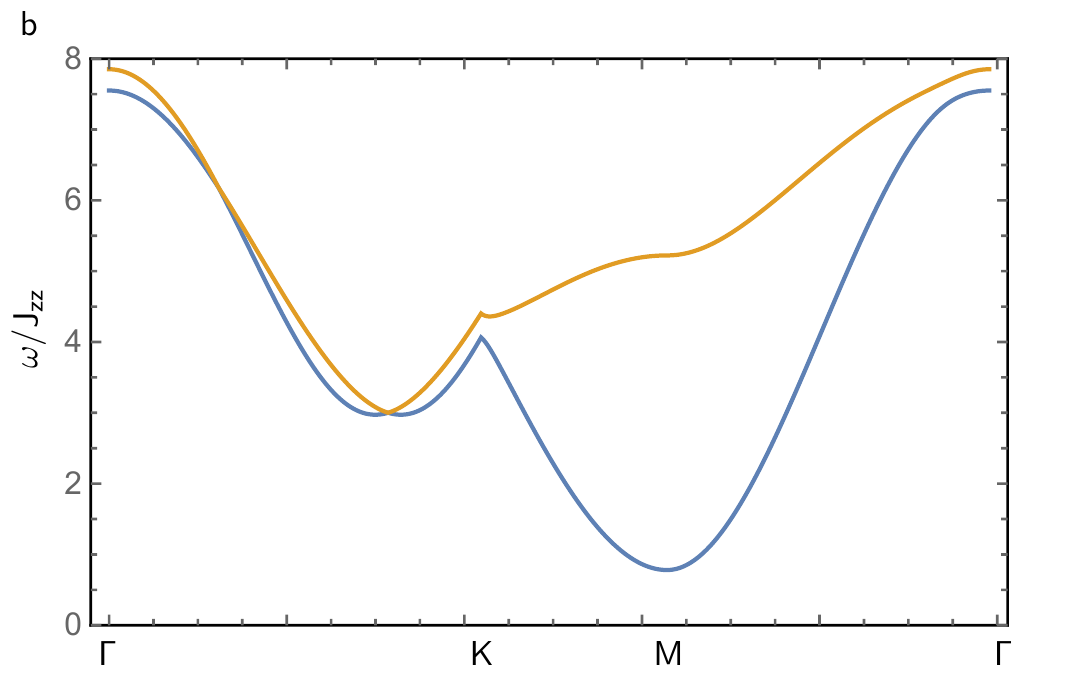}
 \includegraphics[width=8cm]{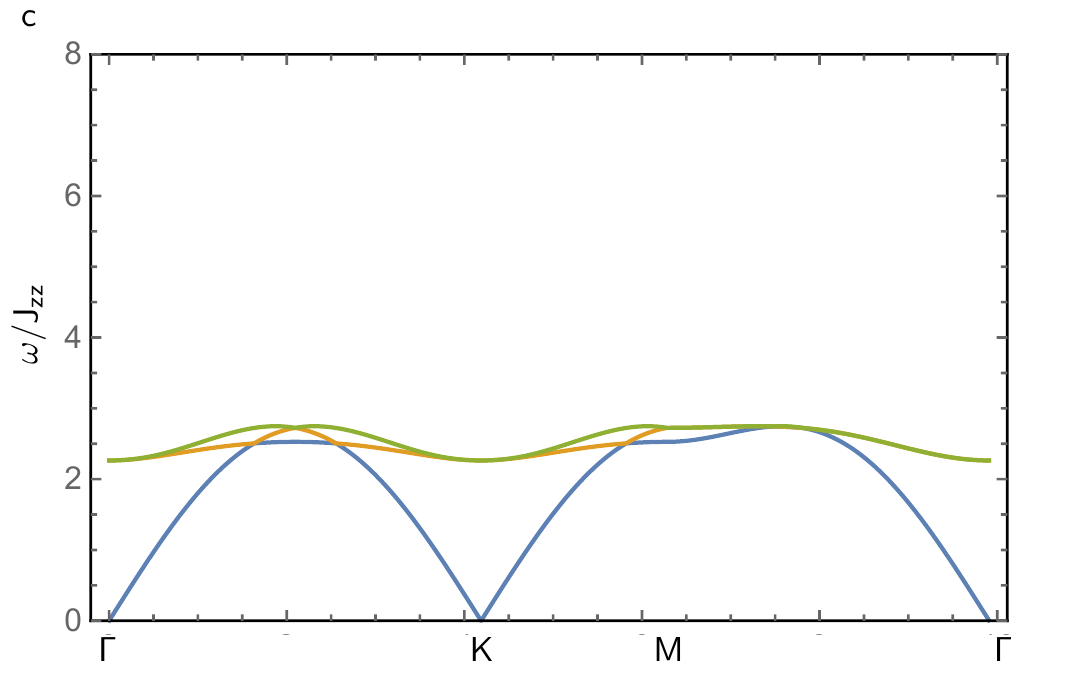}
}
\caption{(Color online.)
Spin wave dispersion along high symmetry momentum points.
(a) Spin wave dispersion in $x$-stripe phase, at $J_{\pm\pm} = -0.9J_{zz}, J_{z\pm} = 0.1 J_{zz}.$ Inset: The first Brillouin zone, with red loop of high-symmetry points along which we plot the dispersion indicated.
(b) Spin wave dispersion in $yz$-stripe phase, at $J_{\pm\pm} = 0.8J_{zz}, J_{z\pm} = 0.8 J_{zz}.$
(c) Spin wave dispersion in $120^\circ$ phase, at $J_{\pm\pm} = J_{z\pm} = 0.$
 }
\label{fig9}
\end{figure}

Since the quantum fluctuation is found to be very weak deep inside each ordered 
phases, it is legitimate to apply the linear spin wave theory to 
study the magnetic excitation in the strongly ordered regimes. 
In Fig.~\ref{fig9}, we plot the representative spin wave dispersions for the 
three ordered phases. Due to the anisotropic spin interaction,
the system does not have any continuous symmetry, so generically the spin wave spectrum
is fully gapped. This is indeed the case for the two stripe ordered phase in Fig.~\ref{fig9}a,b. 
In Fig.~\ref{fig9}c, the parameters are chosen that the spin model reduces to 
a XXZ model. Due to the continuous U(1) symmetry breaking, the spin wave spectrum 
has one gapless mode. As one moves away from this special point, we expect the spectrum should be gapped.

\subsection{Polarized phases and strong magnetic fields}

\begin{figure}[t]
{
 \includegraphics[width=8cm]{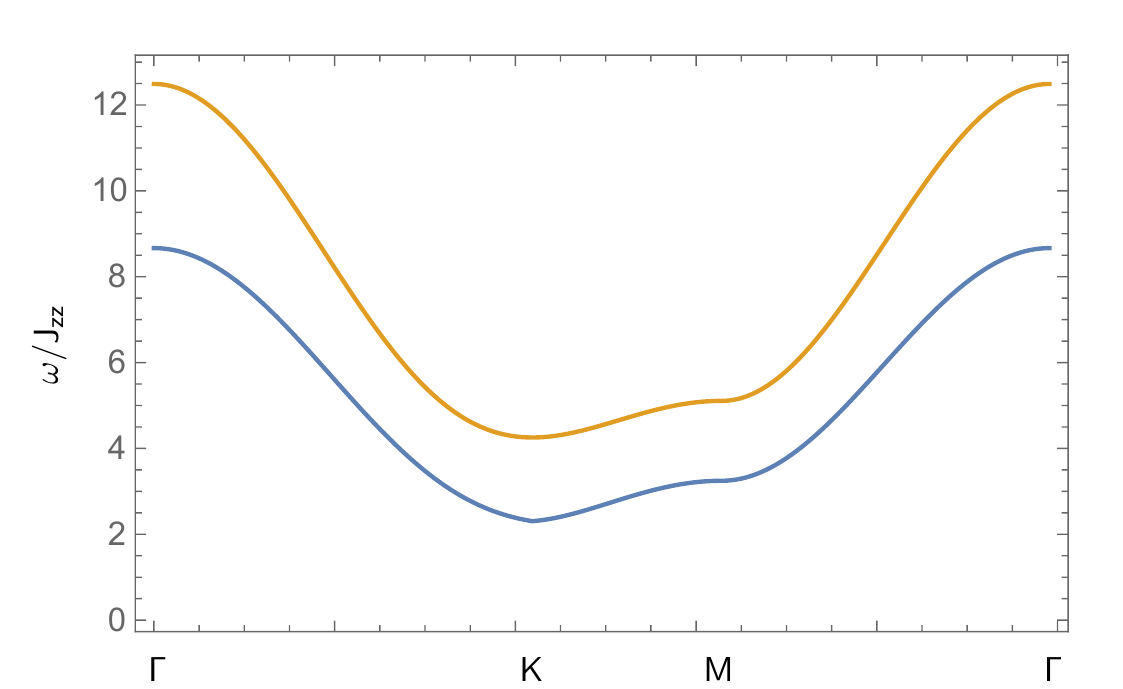}
}
\caption{(Color online.)
Spin wave dispersion when applying external magnetic field along $x$-direction (upper yellow line) and $z$-direction (lower blue line). Here the external field $h$ is taken to be $10J_{zz}$, and anisotropic exchange couplings $J_{z\pm}$ and $J_{\pm\pm}$ are taken to be $0.3J_{zz}$ and $0.2J_{zz}$, respectively.
}
\label{fig10}
\end{figure}

For the rare earth magnets, the $4f$ electrons are very localized. 
As a result, the exchange interaction between the rare earth local 
moments are usually very small. For YbMgGaO$_4$,
the couplings in the spin Hamiltonian are of the order of 1-4K. Therefore, 
an external magnetic field of the order of $10$T is probably 
sufficient to polarize the local moments. 
For the magnetic field that is applied along the $z$ direction, 
we have the spin Hamiltonian,
\begin{equation}
H_{\text Z} = H - h \sum_i S^z_i,
\end{equation}
When the field $h$ is strong enough, the spin 
is polarized along $z$. To obtain the magnetic 
excitation of this polarized state, we use the 
linear spin wave theory and transform the spin operators as
\begin{eqnarray}
S^z_i   & = & \frac{1}{2} - {c}^{\dagger}_i c^{\phantom\dagger}_i , \\
S^{+}_i & = & c^{\phantom\dagger}_i , \\
S^-_i   & = & c^\dagger_i . 
\end{eqnarray} 
We then plug this transformation in the Hamiltonian $H_{\text Z}$
and keep the bilinear terms of boson operators. 
The magnetic excitation only has one branch and is simply given by
\begin{eqnarray}
\Omega_{\text{Z},{\bf k}} & = & \{ [ h - 3 J_{zz} 
+ 2 J_{\pm} \sum_{i=1}^3 \cos ( {\bf k}\cdot {\bf a}_i) ]^2
\nonumber \\
&&  - 4 J_{\pm\pm}^2 | \cos ( {\bf k}\cdot {\bf a}_1) 
+ e^{-i \frac{2\pi}{3} }\cos ( {\bf k}\cdot {\bf a}_2)
\nonumber \\
&& 
+ e^{i \frac{2\pi}{3}  }\cos ( {\bf k} \cdot {\bf a}_3) |^2 \}^{1/2}.  
\end{eqnarray}
The $J_{z\pm}$ coupling is absent in the above spin wave dispersion. 
This is because the $J_{z\pm}$ interaction does not generate any 
quadratic term to the spin wave Hamiltonian. 

For the external field in the $x$ direction, we have 
\begin{eqnarray}
H_{\text X} = H - h \sum_i S^x_i. 
\end{eqnarray}
In the strong field limit, the local moment is polarized along $x$ direction, 
and we transform the spin operators as
\begin{eqnarray}
S^x_i &=& \frac{1}{2} - {d}^{\dagger}_i d^{\phantom\dagger}_i , \\
S^y_i &=& \frac{1}{2} (  d^{\phantom\dagger}_i + {d}^{\dagger}_i) , \\
S^z_i &=& \frac{1}{2i} ( d^{\phantom\dagger}_i - {d}^{\dagger}_i).
\end{eqnarray}
Under the linear spin wave approximation, the magnetic excitation 
is given as 
\begin{eqnarray}
\Omega_{\text{X},{\bf k}} & = & \big\{ \big[
( h - 6 J_{\pm} )
+   ( J_{\pm} - J_{\pm\pm} + \frac{J_{zz}}{2} ) 
\cos ( {\bf k}\cdot {\bf a}_1 ) 
\nonumber \\
&&
+ ( J_{\pm} + \frac{ J_{\pm\pm}}{2} + \frac{J_{zz}}{2} ) 
[ \cos ({\bf k}\cdot {\bf a}_2) + 
   \cos ({\bf k}\cdot {\bf a}_3) ] \big]^2
\nonumber \\
&& 
-  | (
J_{\pm} - J_{\pm\pm}- 
       \frac{J_{zz}}{2}  + i J_{z\pm} ) \cos ( {\bf k}\cdot {\bf a}_1 )  
       \nonumber \\
       && 
  + (J_{\pm} + \frac{J_{\pm\pm}}{2}- \frac{J_{zz}}{2} - i \frac{J_{z\pm}}{2} ) 
  \nonumber \\
&&   \times
[\cos ( {\bf k}\cdot {\bf a}_2 ) + \cos ( {\bf k}\cdot {\bf a}_3 )]  
  |^2   
\big\}^{1/2},
\end{eqnarray}
where all the four couplings enter into the dispersion. 
In Fig.~\ref{fig10}, we plot the spin wave dispersion along high symmetry momentum
points. In practice, it is ready to measure the dynamic spin structure factor 
in YbMgGaO$_4$
and other rare-earth triangular antiferromagnets 
to extract the spin wave dispersion in the strong magnetic fields. 
By comparing the dispersion with the theoretical prediction, one may fully specify
the microscopic spin Hamiltonian and quantitatively determine all the couplings.

\section{Discussion}
\label{sec6}

Our initial treatment of the generic spin model in Sec.~\ref{sec3}
is semiclassical, and the classical ground states that we found 
are magnetically ordered. Due to the strong spin anisotropy that 
arises from the strong SOC, the orientation of the 
local moments is locked with the ordering wavevector in these classical 
orders.  The magnetic excitations in different ordered phases generically
have an excitation gap. Again, this is the spin anisotropy 
that completely breaks all the continuous spin rotational symmetry. 

Our results are quite suggestive for identifying the parameter 
regime of possible disordered ground state of the generic 
model for the quantum spins. 
As we show in Sec.~\ref{sec4}, the 
quantum fluctuation is indeed quite strong in certain parameter regime 
and could in fact completely destroy the magnetic order.  
Therefore, if a quantum spin liquid state does appear in 
the phase diagram of the generic spin model for the triangular lattice, 
it would most likely occur in these frustrated regions near 
the phase boundaries between different ordered states.

\subsection{Materials survey}

Here we turn to a discussion of relevant materials that
have been studied experimentally. 

\subsubsection{YbMgGaO$_4$}

The Yb$^{3+}$ local moments in YbMgGaO$_4$ were found to be disordered  
down to the lowest measurable temperature in the existing experiments. 
It was suggested to be a U(1) quantum spin liquid with a 
spinon Fermi surface~\cite{Lee05}
by one of the author and collaborators~\cite{Yuesheng2015}. 
Whether it is a quantum spin liquid or not is not quite clear at this stage. 
To elucidate the nature of the disordered ground state, an inelastic neutron 
scattering measurement at low temperatures is certainly more desirable. 
On the theoretical side, however, it is more helpful to know precisely 
the actual parameters in the generic model for YbMgGaO$_4$. 
The parameters that were determined from the early thermodynamic measurements
do overlap significantly with the disordered parameter region in Fig.~\ref{fig8}. 
As we discuss in Sec.~\ref{sec5}, in the future experiment 
one could apply strong magnetic fields to polarize 
the spin and measure the spin wave dispersion in an 
inelastic neutron scattering measurement. 
After the disordered parameter region and the actual 
parameters are determined, 
the numerical approaches such as variational wavefunction 
and density matrix renormalization group
may be applied.

\subsubsection{RZn$_3$P$_3$, RCd$_3$P$_3$, RZn$_3$As$_3$ and RCd$_3$As$_3$}

In the RZn$_3$P$_3$, RCd$_3$P$_3$, RZn$_3$As$_3$ and RCd$_3$As$_3$ 
materials' family, the rare earth ions, R$^{3+}$, form triangular layers~\cite{Nientiedt1999,Yamada2010,Stoyko2011,1604}. 
Since the interlayer separation is much larger than the intralayer 
lattice constant, one can safely neglect the interlayer coupling.  
In CeCd$_3$P$_3$, the intralayer lattice constant is 4.28$\AA$
while the interlayer distance is 10.5$\AA$~\cite{1604}. 
As we show in Table~\ref{tab1}, almost all compounds in the RZn$_3$P$_3$, 
RCd$_3$P$_3$, RZn$_3$As$_3$ and RCd$_3$As$_3$ family have the space group 
P$6_3/mmc$. After we restrict the space group symmetry of P$6_3/mmc$
to a single triangular layer, the remaining symmetry elements are 
identical to the ones that are listed for YbMgGaO$_4$ in Sec.~\ref{sec2}. 
Therefore, if the rare-earth local moment in this family of materials
is the same kind of Kramers' doublet as the Yb$^{3+}$ ion in YbMgGaO$_4$,
the local moment interaction is described by the same generic model 
in Eq.~(\ref{eq1}). From the crystal electric field analysis in
Ref.~\onlinecite{1604}, the ground state doublet of the Ce$^{3+}$ 
ion in CeCd$_3$P$_3$ does belong to the same kind of Kramers' doublet as 
the Yb$^{3+}$ ion in YbMgGaO$_4$. Because the system remains paramagnetic 
down to 0.48K, the authors in Ref.~\onlinecite{1604} proposed a
possible quantum spin liquid ground state. One may thus wonder
whether the possible quantum spin liquid in CeCd$_3$P$_3$
is in the same phase as the one that was proposed for YbMgGaO$_4$.
More experiments such as neutron scattering or NMR measurments 
on single crystal samples are certainly needed. 

Among this family of materials, CeZn$_3$P$_3$ is known to develop an
antiferromagnetic order at 0.8K~\cite{Yamada2010}. The precise magnetic ordering 
structure in not known from the existing experiments. 
It is thus of great interest to examine whether the antiferromagnetic
order in CeZn$_3$P$_3$ belongs to one of the orders in our phase diagram. 

If the rare-earth ion contains even number of $4f$ electrons
like the Pr$^{3+}$ ion in PrZn$_3$As$_3$, the local ground state 
doublet is a non-Kramers' doublet. For such a doublet, the in-plane
components, $S^x$ and $S^y$, of the effective spin are even under
time reversal symmetry, while the out-of-plane component, $S^z$, is
odd under time reversal symmetry. This property immediately forbids 
the presence of the $J_{z\pm}$ term in Eq.~(\ref{eq1}) for the 
local moment interaction. In the simplified model, there are 
only three parameters: $J_{zz}$, $J_{\pm}$ and $J_{\pm\pm}$. 
The analysis of the full phase diagram of the simplified model for 
non-Kramers' doublets will be left for future work. 

Finally, we point out one special Kramers' doublet
that was dubbed dipole-octupole Kramers' doublet in Ref.~\onlinecite{huang2014}.
Dipole-octupole Kramers' doublet occurs when the two ground state wavefunctions of the 
doublet are linear superpositions of states that have $J^z$ equal to odd 
integer multiples of $3/2$. Each ground state corresponds to
the 1-dimensional irreducible representation of the $D_{3d}$ point group. 
The two fold Kramers' degeneracy is protected by time reversal symmetry. 
Dipole-octupole doublet is found to exist in the Nd$^{3+}$ ions of 
various Nd-based pyrochlore materials and the Ce$^{3+}$ ion of Ce$_2$Sn$_2$O$_7$~\cite{
PhysRevLett.115.097202,huang2014,PhysRevLett.115.197202,PhysRevB.92.144423,
PhysRevB.92.184418,PhysRevB.92.224430,PhysRevB.91.174416}.  
If the R$^{3+}$ ion in some compounds in the triangular antiferromagnets
belongs to dipole-octupole doublet, the local moment interaction
would be described by an XYZ-like model. We will discuss the dipole-octupole doublet 
in a forthcoming work. Nevertheless, the Yb$^{3+}$ ion in YbMgGaO$_4$ and 
the Ce$^{3+}$ ion in CeCd$_3$P$_3$ are not dipole-octupole doublets~\cite{Yueshengscirep2015,1604}.         

\subsubsection{R$_2$O$_2$CO$_3$}

Rare-earth oxy-carbonates R$_2$O$_2$CO$_3$ (R = Nd, Sm, Dy) is another layered triangular antiferromagnet 
family that is discovered quite recently. All these materials crystallize in a hexagonal 
structure with the same space group as CeCd$_3$P$_3$. 
Although all three magnetic ions are Kramers' doublet, 
the crystal field ground states are not carefully studied in the existing experiments~\cite{Arjun2016}. If these Kramers' doublets are not dipole-octupole doublets,
the local moment interaction is given by our anisotropic model in Eq.~\ref{eq1}. 
The precise magnetic ordered structure is unknown, so it is of interest
to study the magnetic structure and excitation of these systems.

\subsection{Summary}

To summarize, we analyzed the generic spin 
Hamiltonian that describes the interaction between the spin-orbit-entangled 
Kramers' doublet local moments. We have obtained the magnetic 
phase diagram that includes three distinct ordered phase. We 
have further identified the possible disordered region 
that might host disordered ground states 
for the Yb local moments in YbMgGaO$_4$~\cite{Yuesheng2015}. 
We carefully studied the magnetic excitation in different ordered
phases as well as in the presence of strong magnetic fields. 
The ordered phases and the magnetic excitations
may be detected in the future experiments in
the strong spin-orbit-coupled triangular antiferromagnets.

As the generic model applies to any other Kramers' doublet with the same
symmetry properties, to further justify the 
applicability of this model, it is thus of great interest to experimentally
study the magnetic properties of other rare-earth based triangular materials
and access the magnetic orders in the phase diagram 
and the magnetic transition from the possible disordered states.
Apart from the rare-earth systems, the spin-orbit-entangled Kramers' 
doublet local moments can appear in the partially filled 
$t_{2g}$ shells such as triangular lattice iridates~\cite{WCKB}. 
Given the $4d$ or $5d$ nature of the local moments,
the exchange interaction is certainly enhanced, and thus one could probe 
the magnetic properties at a much higher temperature than the rare-earth systems. 

\vspace{3mm}

\section{Acknowledgements}

We acknowledge Yuesheng Li (Renmin Univ) and Qingming Zhang (Renmin Univ) for 
the previous collaboration, Leon Balents and Patrick Lee for correspondences.
We thank Xi Dai, Fengren Fan, Jiangping Hu, Dunghai Lee, Shiyan Li, 
Yuanming Lu, Yang Qi, Fa Wang, Nanlin Wang, Zhong Wang, Hua Wu, Hong Yao,  
Rong Yu, Yue Yu, Jize Zhao, Fuchun Zhang, Jun Zhao, 
and Yi Zhou for valuable conversation. 
We are particularly indebted to Zhong Wang at 
Institute of Advanced Study of Tsinghua University
for hosting our stay where part of the work was completed. 
This work is supported by the start-up fund of Fudan University
and the National Thousand-Young-Talents Program. 

\bibliography{refs}

\begin{thebibliography}{41}%
\makeatletter
\providecommand \@ifxundefined [1]{%
 \@ifx{#1\undefined}
}%
\providecommand \@ifnum [1]{%
 \ifnum #1\expandafter \@firstoftwo
 \else \expandafter \@secondoftwo
 \fi
}%
\providecommand \@ifx [1]{%
 \ifx #1\expandafter \@firstoftwo
 \else \expandafter \@secondoftwo
 \fi
}%
\providecommand \natexlab [1]{#1}%
\providecommand \enquote  [1]{``#1''}%
\providecommand \bibnamefont  [1]{#1}%
\providecommand \bibfnamefont [1]{#1}%
\providecommand \citenamefont [1]{#1}%
\providecommand \href@noop [0]{\@secondoftwo}%
\providecommand \href [0]{\begingroup \@sanitize@url \@href}%
\providecommand \@href[1]{\@@startlink{#1}\@@href}%
\providecommand \@@href[1]{\endgroup#1\@@endlink}%
\providecommand \@sanitize@url [0]{\catcode `\\12\catcode `\$12\catcode
  `\&12\catcode `\#12\catcode `\^12\catcode `\_12\catcode `\%12\relax}%
\providecommand \@@startlink[1]{}%
\providecommand \@@endlink[0]{}%
\providecommand \url  [0]{\begingroup\@sanitize@url \@url }%
\providecommand \@url [1]{\endgroup\@href {#1}{\urlprefix }}%
\providecommand \urlprefix  [0]{URL }%
\providecommand \Eprint [0]{\href }%
\providecommand \doibase [0]{http://dx.doi.org/}%
\providecommand \selectlanguage [0]{\@gobble}%
\providecommand \bibinfo  [0]{\@secondoftwo}%
\providecommand \bibfield  [0]{\@secondoftwo}%
\providecommand \translation [1]{[#1]}%
\providecommand \BibitemOpen [0]{}%
\providecommand \bibitemStop [0]{}%
\providecommand \bibitemNoStop [0]{.\EOS\space}%
\providecommand \EOS [0]{\spacefactor3000\relax}%
\providecommand \BibitemShut  [1]{\csname bibitem#1\endcsname}%
\let\auto@bib@innerbib\@empty
\bibitem [{\citenamefont {Hasan}\ and\ \citenamefont
  {Kane}(2010)}]{KaneReview}%
  \BibitemOpen
  \bibfield  {author} {\bibinfo {author} {\bibfnamefont {M.~Z.}\ \bibnamefont
  {Hasan}}\ and\ \bibinfo {author} {\bibfnamefont {C.~L.}\ \bibnamefont
  {Kane}},\ }\bibfield  {title} {\enquote {\bibinfo {title}
  {\textit{Colloquium} : Topological insulators},}\ }\href {\doibase
  10.1103/RevModPhys.82.3045} {\bibfield  {journal} {\bibinfo  {journal} {Rev.
  Mod. Phys.}\ }\textbf {\bibinfo {volume} {82}},\ \bibinfo {pages}
  {3045--3067} (\bibinfo {year} {2010})}\BibitemShut {NoStop}%
\bibitem [{\citenamefont {Witczak-Krempa}\ \emph {et~al.}(2014)\citenamefont
  {Witczak-Krempa}, \citenamefont {Chen}, \citenamefont {Kim},\ and\
  \citenamefont {Balents}}]{WCKB}%
  \BibitemOpen
  \bibfield  {author} {\bibinfo {author} {\bibfnamefont {William}\ \bibnamefont
  {Witczak-Krempa}}, \bibinfo {author} {\bibfnamefont {Gang}\ \bibnamefont
  {Chen}}, \bibinfo {author} {\bibfnamefont {Yong~Baek}\ \bibnamefont {Kim}}, \
  and\ \bibinfo {author} {\bibfnamefont {Leon}\ \bibnamefont {Balents}},\
  }\bibfield  {title} {\enquote {\bibinfo {title} {Correlated quantum phenomena
  in the strong spin-orbit regime},}\ }\href@noop {} {\bibfield  {journal}
  {\bibinfo  {journal} {Annual Review of Condensed Matter Physics}\ }\textbf
  {\bibinfo {volume} {5}},\ \bibinfo {pages} {57--82} (\bibinfo {year}
  {2014})}\BibitemShut {NoStop}%
\bibitem [{\citenamefont {Li}\ \emph {et~al.}(2015{\natexlab{a}})\citenamefont
  {Li}, \citenamefont {Chen}, \citenamefont {Tong}, \citenamefont {Pi},
  \citenamefont {Liu}, \citenamefont {Yang}, \citenamefont {Wang},\ and\
  \citenamefont {Zhang}}]{Yuesheng2015}%
  \BibitemOpen
  \bibfield  {author} {\bibinfo {author} {\bibfnamefont {Yuesheng}\
  \bibnamefont {Li}}, \bibinfo {author} {\bibfnamefont {Gang}\ \bibnamefont
  {Chen}}, \bibinfo {author} {\bibfnamefont {Wei}\ \bibnamefont {Tong}},
  \bibinfo {author} {\bibfnamefont {Li}~\bibnamefont {Pi}}, \bibinfo {author}
  {\bibfnamefont {Juanjuan}\ \bibnamefont {Liu}}, \bibinfo {author}
  {\bibfnamefont {Zhaorong}\ \bibnamefont {Yang}}, \bibinfo {author}
  {\bibfnamefont {Xiaoqun}\ \bibnamefont {Wang}}, \ and\ \bibinfo {author}
  {\bibfnamefont {Qingming}\ \bibnamefont {Zhang}},\ }\bibfield  {title}
  {\enquote {\bibinfo {title} {Rare-earth triangular lattice spin liquid: A
  single-crystal study of ${\mathrm{ybmggao}}_{4}$},}\ }\href {\doibase
  10.1103/PhysRevLett.115.167203} {\bibfield  {journal} {\bibinfo  {journal}
  {Phys. Rev. Lett.}\ }\textbf {\bibinfo {volume} {115}},\ \bibinfo {pages}
  {167203} (\bibinfo {year} {2015}{\natexlab{a}})}\BibitemShut {NoStop}%
\bibitem [{\citenamefont {Li}\ \emph {et~al.}(2015{\natexlab{b}})\citenamefont
  {Li}, \citenamefont {Liao}, \citenamefont {Zhang}, \citenamefont {Li},
  \citenamefont {Jin}, \citenamefont {Ling}, \citenamefont {Zhang},
  \citenamefont {Zou}, \citenamefont {Pi}, \citenamefont {Yang}, \citenamefont
  {Wang}, \citenamefont {Wu},\ and\ \citenamefont
  {Zhang}}]{Yueshengscirep2015}%
  \BibitemOpen
  \bibfield  {author} {\bibinfo {author} {\bibfnamefont {Yuesheng}\
  \bibnamefont {Li}}, \bibinfo {author} {\bibfnamefont {Haijun}\ \bibnamefont
  {Liao}}, \bibinfo {author} {\bibfnamefont {Zhen}\ \bibnamefont {Zhang}},
  \bibinfo {author} {\bibfnamefont {Shiyan}\ \bibnamefont {Li}}, \bibinfo
  {author} {\bibfnamefont {Feng}\ \bibnamefont {Jin}}, \bibinfo {author}
  {\bibfnamefont {Langsheng}\ \bibnamefont {Ling}}, \bibinfo {author}
  {\bibfnamefont {Lei}\ \bibnamefont {Zhang}}, \bibinfo {author} {\bibfnamefont
  {Youming}\ \bibnamefont {Zou}}, \bibinfo {author} {\bibfnamefont
  {Li}~\bibnamefont {Pi}}, \bibinfo {author} {\bibfnamefont {Zhaorong}\
  \bibnamefont {Yang}}, \bibinfo {author} {\bibfnamefont {Junfeng}\
  \bibnamefont {Wang}}, \bibinfo {author} {\bibfnamefont {Zhonghua}\
  \bibnamefont {Wu}}, \ and\ \bibinfo {author} {\bibfnamefont {Qingming}\
  \bibnamefont {Zhang}},\ }\bibfield  {title} {\enquote {\bibinfo {title}
  {Gapless quantum spin liquid ground state in the two-dimensional spin-1/2
  triangular antiferromagnet ybmggao4},}\ }\href {\doibase 10.1038/srep16419}
  {\bibfield  {journal} {\bibinfo  {journal} {Scientific Reports}\ }\textbf
  {\bibinfo {volume} {5}},\ \bibinfo {pages} {16419} (\bibinfo {year}
  {2015}{\natexlab{b}})}\BibitemShut {NoStop}%
\bibitem [{\citenamefont {Higuchi}\ \emph {et~al.}(2016)\citenamefont
  {Higuchi}, \citenamefont {Noshima}, \citenamefont {Shirakawa}, \citenamefont
  {Tsubota}, ,\ and\ \citenamefont {Kitagawa}}]{1604}%
  \BibitemOpen
  \bibfield  {author} {\bibinfo {author} {\bibfnamefont {S.}~\bibnamefont
  {Higuchi}}, \bibinfo {author} {\bibfnamefont {Y.}~\bibnamefont {Noshima}},
  \bibinfo {author} {\bibfnamefont {N.}~\bibnamefont {Shirakawa}}, \bibinfo
  {author} {\bibfnamefont {M.}~\bibnamefont {Tsubota}}, , \ and\ \bibinfo
  {author} {\bibfnamefont {J.}~\bibnamefont {Kitagawa}},\ }\bibfield  {title}
  {\enquote {\bibinfo {title} {Optical, transport and magnetic properties of
  new compound cecd3p3},}\ }\href@noop {} {\bibfield  {journal} {\bibinfo
  {journal} {arXiv:1604.04016}\ } (\bibinfo {year} {2016})}\BibitemShut
  {NoStop}%
\bibitem [{\citenamefont {Nientiedt}\ and\ \citenamefont
  {Jeitschko}(1999)}]{Nientiedt1999}%
  \BibitemOpen
  \bibfield  {author} {\bibinfo {author} {\bibfnamefont {A.T.}\ \bibnamefont
  {Nientiedt}}\ and\ \bibinfo {author} {\bibfnamefont {W.}~\bibnamefont
  {Jeitschko}},\ }\bibfield  {title} {\enquote {\bibinfo {title} {The series of
  rare earth zinc phosphides rzn3p3 (r=y, la-nd, sm, gd-er) and the
  corresponding cadmium compound prcd3p3},}\ }\href@noop {} {\bibfield
  {journal} {\bibinfo  {journal} {Journal of Solid State Chemistry}\ }\textbf
  {\bibinfo {volume} {146}},\ \bibinfo {pages} {483} (\bibinfo {year}
  {1999})}\BibitemShut {NoStop}%
\bibitem [{\citenamefont {Yamada}\ \emph {et~al.}(2010)\citenamefont {Yamada},
  \citenamefont {Hara}, \citenamefont {Matsubayashi}, \citenamefont {Munakata},
  \citenamefont {Ganguli}, \citenamefont {Ochiai}, \citenamefont {Matsumoto},\
  and\ \citenamefont {Uwatoko}}]{Yamada2010}%
  \BibitemOpen
  \bibfield  {author} {\bibinfo {author} {\bibfnamefont {A.}~\bibnamefont
  {Yamada}}, \bibinfo {author} {\bibfnamefont {N.}~\bibnamefont {Hara}},
  \bibinfo {author} {\bibfnamefont {K.}~\bibnamefont {Matsubayashi}}, \bibinfo
  {author} {\bibfnamefont {K.}~\bibnamefont {Munakata}}, \bibinfo {author}
  {\bibfnamefont {C.}~\bibnamefont {Ganguli}}, \bibinfo {author} {\bibfnamefont
  {A.}~\bibnamefont {Ochiai}}, \bibinfo {author} {\bibfnamefont
  {T.}~\bibnamefont {Matsumoto}}, \ and\ \bibinfo {author} {\bibfnamefont
  {Y.}~\bibnamefont {Uwatoko}},\ }\bibfield  {title} {\enquote {\bibinfo
  {title} {Effect of pressure on the electrical resistivity of cezn3p3},}\
  }\href {\doibase 10.1088/1742-6596/215/1/012031} {\bibfield  {journal}
  {\bibinfo  {journal} {J. Phys.: Conf. Ser.}\ }\textbf {\bibinfo {volume}
  {215}},\ \bibinfo {pages} {012031} (\bibinfo {year} {2010})}\BibitemShut
  {NoStop}%
\bibitem [{\citenamefont {Stoyko}\ and\ \citenamefont
  {Mar}(2011)}]{Stoyko2011}%
  \BibitemOpen
  \bibfield  {author} {\bibinfo {author} {\bibfnamefont {Stanislav~S.}\
  \bibnamefont {Stoyko}}\ and\ \bibinfo {author} {\bibfnamefont {Arthur}\
  \bibnamefont {Mar}},\ }\bibfield  {title} {\enquote {\bibinfo {title}
  {Ternary rare-earth arsenides rezn3as3 (re = la-nd, sm) and recd3as3 (re =
  la-pr)},}\ }\href {\doibase dx.doi.org/10.1021/ic201708x} {\bibfield
  {journal} {\bibinfo  {journal} {Inorg. Chem}\ }\textbf {\bibinfo {volume}
  {50}},\ \bibinfo {pages} {11152–11161} (\bibinfo {year}
  {2011})}\BibitemShut {NoStop}%
\bibitem [{\citenamefont {Arjun}\ \emph {et~al.}(2016)\citenamefont {Arjun},
  \citenamefont {Brinda}, \citenamefont {Padmanabhan},\ and\ \citenamefont
  {Nath}}]{Arjun2016}%
  \BibitemOpen
  \bibfield  {author} {\bibinfo {author} {\bibfnamefont {U.}~\bibnamefont
  {Arjun}}, \bibinfo {author} {\bibfnamefont {K.}~\bibnamefont {Brinda}},
  \bibinfo {author} {\bibfnamefont {M.}~\bibnamefont {Padmanabhan}}, \ and\
  \bibinfo {author} {\bibfnamefont {R.}~\bibnamefont {Nath}},\ }\bibfield
  {title} {\enquote {\bibinfo {title} {Magnetic properties of layered
  rare-earth oxy-carbonates ln2o2co3 (ln=nd, sm, and dy)},}\ }\href {\doibase
  http://dx.doi.org/10.1016/j.ssc.2016.04.007} {\bibfield  {journal} {\bibinfo
  {journal} {Solid State Communications}\ ,\ \bibinfo {pages} {--}} (\bibinfo
  {year} {2016})}\BibitemShut {NoStop}%
\bibitem [{\citenamefont {Chen}\ and\ \citenamefont
  {Balents}(2008)}]{Chen2008}%
  \BibitemOpen
  \bibfield  {author} {\bibinfo {author} {\bibfnamefont {Gang}\ \bibnamefont
  {Chen}}\ and\ \bibinfo {author} {\bibfnamefont {Leon}\ \bibnamefont
  {Balents}},\ }\bibfield  {title} {\enquote {\bibinfo {title} {Spin-orbit
  effects in ${\text{na}}_{4}{\text{ir}}_{3}{\text{o}}_{8}$: A hyper-kagome
  lattice antiferromagnet},}\ }\href {\doibase 10.1103/PhysRevB.78.094403}
  {\bibfield  {journal} {\bibinfo  {journal} {Phys. Rev. B}\ }\textbf {\bibinfo
  {volume} {78}},\ \bibinfo {pages} {094403} (\bibinfo {year}
  {2008})}\BibitemShut {NoStop}%
\bibitem [{\citenamefont {Li}\ \emph {et~al.}(2015{\natexlab{c}})\citenamefont
  {Li}, \citenamefont {Yu},\ and\ \citenamefont {Li}}]{JXLi2015}%
  \BibitemOpen
  \bibfield  {author} {\bibinfo {author} {\bibfnamefont {Kai}\ \bibnamefont
  {Li}}, \bibinfo {author} {\bibfnamefont {Shun-Li}\ \bibnamefont {Yu}}, \ and\
  \bibinfo {author} {\bibfnamefont {Jian-Xin}\ \bibnamefont {Li}},\ }\bibfield
  {title} {\enquote {\bibinfo {title} {Global phase diagram, possible chiral
  spin liquid, and topological superconductivity in the triangular
  kitaev–heisenberg model},}\ }\href
  {http://stacks.iop.org/1367-2630/17/i=4/a=043032} {\bibfield  {journal}
  {\bibinfo  {journal} {New Journal of Physics}\ }\textbf {\bibinfo {volume}
  {17}},\ \bibinfo {pages} {043032} (\bibinfo {year}
  {2015}{\natexlab{c}})}\BibitemShut {NoStop}%
\bibitem [{\citenamefont {Chen}\ \emph {et~al.}(2010)\citenamefont {Chen},
  \citenamefont {Pereira},\ and\ \citenamefont {Balents}}]{Chen2010}%
  \BibitemOpen
  \bibfield  {author} {\bibinfo {author} {\bibfnamefont {Gang}\ \bibnamefont
  {Chen}}, \bibinfo {author} {\bibfnamefont {Rodrigo}\ \bibnamefont {Pereira}},
  \ and\ \bibinfo {author} {\bibfnamefont {Leon}\ \bibnamefont {Balents}},\
  }\bibfield  {title} {\enquote {\bibinfo {title} {Exotic phases induced by
  strong spin-orbit coupling in ordered double perovskites},}\ }\href {\doibase
  10.1103/PhysRevB.82.174440} {\bibfield  {journal} {\bibinfo  {journal} {Phys.
  Rev. B}\ }\textbf {\bibinfo {volume} {82}},\ \bibinfo {pages} {174440}
  (\bibinfo {year} {2010})}\BibitemShut {NoStop}%
\bibitem [{\citenamefont {Curnoe}(2008)}]{Curnoe2008}%
  \BibitemOpen
  \bibfield  {author} {\bibinfo {author} {\bibfnamefont {S.~H.}\ \bibnamefont
  {Curnoe}},\ }\bibfield  {title} {\enquote {\bibinfo {title} {Structural
  distortion and the spin liquid state in
  ${\text{tb}}_{2}{\text{ti}}_{2}{\text{o}}_{7}$},}\ }\href {\doibase
  10.1103/PhysRevB.78.094418} {\bibfield  {journal} {\bibinfo  {journal} {Phys.
  Rev. B}\ }\textbf {\bibinfo {volume} {78}},\ \bibinfo {pages} {094418}
  (\bibinfo {year} {2008})}\BibitemShut {NoStop}%
\bibitem [{\citenamefont {Onoda}\ and\ \citenamefont
  {Tanaka}(2010)}]{Onoda2010}%
  \BibitemOpen
  \bibfield  {author} {\bibinfo {author} {\bibfnamefont {Shigeki}\ \bibnamefont
  {Onoda}}\ and\ \bibinfo {author} {\bibfnamefont {Yoichi}\ \bibnamefont
  {Tanaka}},\ }\bibfield  {title} {\enquote {\bibinfo {title} {Quantum melting
  of spin ice: Emergent cooperative quadrupole and chirality},}\ }\href
  {\doibase 10.1103/PhysRevLett.105.047201} {\bibfield  {journal} {\bibinfo
  {journal} {Phys. Rev. Lett.}\ }\textbf {\bibinfo {volume} {105}},\ \bibinfo
  {pages} {047201} (\bibinfo {year} {2010})}\BibitemShut {NoStop}%
\bibitem [{\citenamefont {Chen}\ and\ \citenamefont
  {Balents}(2011)}]{Chen2011}%
  \BibitemOpen
  \bibfield  {author} {\bibinfo {author} {\bibfnamefont {Gang}\ \bibnamefont
  {Chen}}\ and\ \bibinfo {author} {\bibfnamefont {Leon}\ \bibnamefont
  {Balents}},\ }\bibfield  {title} {\enquote {\bibinfo {title} {Spin-orbit
  coupling in ${d}^{2}$ ordered double perovskites},}\ }\href {\doibase
  10.1103/PhysRevB.84.094420} {\bibfield  {journal} {\bibinfo  {journal} {Phys.
  Rev. B}\ }\textbf {\bibinfo {volume} {84}},\ \bibinfo {pages} {094420}
  (\bibinfo {year} {2011})}\BibitemShut {NoStop}%
\bibitem [{\citenamefont {Jackeli}\ and\ \citenamefont
  {Khaliullin}(2009)}]{Jackeli2009}%
  \BibitemOpen
  \bibfield  {author} {\bibinfo {author} {\bibfnamefont {G.}~\bibnamefont
  {Jackeli}}\ and\ \bibinfo {author} {\bibfnamefont {G.}~\bibnamefont
  {Khaliullin}},\ }\bibfield  {title} {\enquote {\bibinfo {title} {Mott
  insulators in the strong spin-orbit coupling limit: From heisenberg to a
  quantum compass and kitaev models},}\ }\href {\doibase
  10.1103/PhysRevLett.102.017205} {\bibfield  {journal} {\bibinfo  {journal}
  {Phys. Rev. Lett.}\ }\textbf {\bibinfo {volume} {102}},\ \bibinfo {pages}
  {017205} (\bibinfo {year} {2009})}\BibitemShut {NoStop}%
\bibitem [{\citenamefont {Ross}\ \emph {et~al.}(2011)\citenamefont {Ross},
  \citenamefont {Savary}, \citenamefont {Gaulin},\ and\ \citenamefont
  {Balents}}]{Ross2011}%
  \BibitemOpen
  \bibfield  {author} {\bibinfo {author} {\bibfnamefont {Kate~A.}\ \bibnamefont
  {Ross}}, \bibinfo {author} {\bibfnamefont {Lucile}\ \bibnamefont {Savary}},
  \bibinfo {author} {\bibfnamefont {Bruce~D.}\ \bibnamefont {Gaulin}}, \ and\
  \bibinfo {author} {\bibfnamefont {Leon}\ \bibnamefont {Balents}},\ }\bibfield
   {title} {\enquote {\bibinfo {title} {Quantum excitations in quantum spin
  ice},}\ }\href {\doibase 10.1103/PhysRevX.1.021002} {\bibfield  {journal}
  {\bibinfo  {journal} {Phys. Rev. X}\ }\textbf {\bibinfo {volume} {1}},\
  \bibinfo {pages} {021002} (\bibinfo {year} {2011})}\BibitemShut {NoStop}%
\bibitem [{\citenamefont {Luttinger}\ and\ \citenamefont
  {Tisza}(1946)}]{PhysRev.70.954}%
  \BibitemOpen
  \bibfield  {author} {\bibinfo {author} {\bibfnamefont {J.~M.}\ \bibnamefont
  {Luttinger}}\ and\ \bibinfo {author} {\bibfnamefont {L.}~\bibnamefont
  {Tisza}},\ }\bibfield  {title} {\enquote {\bibinfo {title} {Theory of dipole
  interaction in crystals},}\ }\href {\doibase 10.1103/PhysRev.70.954}
  {\bibfield  {journal} {\bibinfo  {journal} {Phys. Rev.}\ }\textbf {\bibinfo
  {volume} {70}},\ \bibinfo {pages} {954--964} (\bibinfo {year}
  {1946})}\BibitemShut {NoStop}%
\bibitem [{\citenamefont {Ishizuka}\ and\ \citenamefont
  {Balents}(2015)}]{PhysRevB.92.020411}%
  \BibitemOpen
  \bibfield  {author} {\bibinfo {author} {\bibfnamefont {Hiroaki}\ \bibnamefont
  {Ishizuka}}\ and\ \bibinfo {author} {\bibfnamefont {Leon}\ \bibnamefont
  {Balents}},\ }\bibfield  {title} {\enquote {\bibinfo {title} {Switching of
  magnetic anisotropy in a fcc antiferromagnet with direction-dependent
  interactions},}\ }\href {\doibase 10.1103/PhysRevB.92.020411} {\bibfield
  {journal} {\bibinfo  {journal} {Phys. Rev. B}\ }\textbf {\bibinfo {volume}
  {92}},\ \bibinfo {pages} {020411} (\bibinfo {year} {2015})}\BibitemShut
  {NoStop}%
\bibitem [{Note1()}]{Note1}%
  \BibitemOpen
  \bibinfo {note} {The actual spin orientation depends on the
  couplings.}\BibitemShut {Stop}%
\bibitem [{\citenamefont {Savary}\ and\ \citenamefont
  {Balents}(2012)}]{Savary12}%
  \BibitemOpen
  \bibfield  {author} {\bibinfo {author} {\bibfnamefont {Lucile}\ \bibnamefont
  {Savary}}\ and\ \bibinfo {author} {\bibfnamefont {Leon}\ \bibnamefont
  {Balents}},\ }\bibfield  {title} {\enquote {\bibinfo {title} {Coulombic
  quantum liquids in spin-1/2 pyrochlores},}\ }\href {\doibase
  10.1103/PhysRevLett.108.037202} {\bibfield  {journal} {\bibinfo  {journal}
  {Phys. Rev. Lett.}\ }\textbf {\bibinfo {volume} {108}},\ \bibinfo {pages}
  {037202} (\bibinfo {year} {2012})}\BibitemShut {NoStop}%
\bibitem [{\citenamefont {Kawamura}\ and\ \citenamefont
  {Miyashita}(1984)}]{MonteCarlo1}%
  \BibitemOpen
  \bibfield  {author} {\bibinfo {author} {\bibfnamefont {Hikaru}\ \bibnamefont
  {Kawamura}}\ and\ \bibinfo {author} {\bibfnamefont {Seiji}\ \bibnamefont
  {Miyashita}},\ }\bibfield  {title} {\enquote {\bibinfo {title} {Phase
  transition of the two-dimensional heisenberg antiferromagnet on the
  triangular lattice},}\ }\href {\doibase 10.1143/JPSJ.53.4138} {\bibfield
  {journal} {\bibinfo  {journal} {Journal of the Physical Society of Japan}\
  }\textbf {\bibinfo {volume} {53}},\ \bibinfo {pages} {4138--4154} (\bibinfo
  {year} {1984})}\BibitemShut {NoStop}%
\bibitem [{\citenamefont {Southern}\ and\ \citenamefont
  {Xu}(1995)}]{MonteCarlo2}%
  \BibitemOpen
  \bibfield  {author} {\bibinfo {author} {\bibfnamefont {B.W.}\ \bibnamefont
  {Southern}}\ and\ \bibinfo {author} {\bibfnamefont {H-J.}\ \bibnamefont
  {Xu}},\ }\bibfield  {title} {\enquote {\bibinfo {title} {Monte carlo study of
  the heisenberg antiferromagnet on the triangular lattice},}\ }\href {\doibase
  10.1103/PhysRevB.52.R3836} {\bibfield  {journal} {\bibinfo  {journal} {Phys.
  Rev. B}\ }\textbf {\bibinfo {volume} {52}},\ \bibinfo {pages} {R3836--R3839}
  (\bibinfo {year} {1995})}\BibitemShut {NoStop}%
\bibitem [{\citenamefont {Metropolis}\ \emph {et~al.}(1953)\citenamefont
  {Metropolis}, \citenamefont {Rosenbluth}, \citenamefont {Rosenbluth},
  \citenamefont {Teller},\ and\ \citenamefont {Teller}}]{Metropolis1}%
  \BibitemOpen
  \bibfield  {author} {\bibinfo {author} {\bibfnamefont {Nicholas}\
  \bibnamefont {Metropolis}}, \bibinfo {author} {\bibfnamefont {Arianna~W.}\
  \bibnamefont {Rosenbluth}}, \bibinfo {author} {\bibfnamefont {Marshall~N.}\
  \bibnamefont {Rosenbluth}}, \bibinfo {author} {\bibfnamefont {Augusta~H.}\
  \bibnamefont {Teller}}, \ and\ \bibinfo {author} {\bibfnamefont {Edward}\
  \bibnamefont {Teller}},\ }\bibfield  {title} {\enquote {\bibinfo {title}
  {Equation of state calculations by fast computing machines},}\ }\href
  {\doibase http://dx.doi.org/10.1063/1.1699114} {\bibfield  {journal}
  {\bibinfo  {journal} {The Journal of Chemical Physics}\ }\textbf {\bibinfo
  {volume} {21}},\ \bibinfo {pages} {1087--1092} (\bibinfo {year}
  {1953})}\BibitemShut {NoStop}%
\bibitem [{\citenamefont {Hastings}(1970)}]{Metropolis2}%
  \BibitemOpen
  \bibfield  {author} {\bibinfo {author} {\bibfnamefont {W.~K.}\ \bibnamefont
  {Hastings}},\ }\bibfield  {title} {\enquote {\bibinfo {title} {Monte carlo
  sampling methods using markov chains and their applications},}\ }\href
  {\doibase 10.1093/biomet/57.1.97} {\bibfield  {journal} {\bibinfo  {journal}
  {Biometrika}\ }\textbf {\bibinfo {volume} {57}},\ \bibinfo {pages} {97--109}
  (\bibinfo {year} {1970})}\BibitemShut {NoStop}%
\bibitem [{\citenamefont {Marsaglia}(1972)}]{SpherePoint}%
  \BibitemOpen
  \bibfield  {author} {\bibinfo {author} {\bibfnamefont {George}\ \bibnamefont
  {Marsaglia}},\ }\bibfield  {title} {\enquote {\bibinfo {title} {Choosing a
  point from the surface of a sphere},}\ }\href {\doibase
  10.1214/aoms/1177692644} {\bibfield  {journal} {\bibinfo  {journal} {Ann.
  Math. Statist.}\ }\textbf {\bibinfo {volume} {43}},\ \bibinfo {pages}
  {645--646} (\bibinfo {year} {1972})}\BibitemShut {NoStop}%
\bibitem [{\citenamefont {Binder}(1981)}]{BinderRatio1}%
  \BibitemOpen
  \bibfield  {author} {\bibinfo {author} {\bibfnamefont {K.}~\bibnamefont
  {Binder}},\ }\bibfield  {title} {\enquote {\bibinfo {title} {Finite size
  scaling analysis of ising model block distribution functions},}\ }\href@noop
  {} {\bibfield  {journal} {\bibinfo  {journal} {Zeitschrift für Physik B
  Condensed Matter}\ }\textbf {\bibinfo {volume} {43}},\ \bibinfo {pages}
  {119--140} (\bibinfo {year} {1981})}\BibitemShut {NoStop}%
\bibitem [{\citenamefont {Binder}(1997)}]{BinderRatio2}%
  \BibitemOpen
  \bibfield  {author} {\bibinfo {author} {\bibfnamefont {K}~\bibnamefont
  {Binder}},\ }\bibfield  {title} {\enquote {\bibinfo {title} {Applications of
  monte carlo methods to statistical physics},}\ }\href
  {http://stacks.iop.org/0034-4885/60/i=5/a=001} {\bibfield  {journal}
  {\bibinfo  {journal} {Reports on Progress in Physics}\ }\textbf {\bibinfo
  {volume} {60}},\ \bibinfo {pages} {487} (\bibinfo {year} {1997})}\BibitemShut
  {NoStop}%
\bibitem [{\citenamefont {Ramirez}(1994)}]{Ramirez1994}%
  \BibitemOpen
  \bibfield  {author} {\bibinfo {author} {\bibfnamefont {A~P}\ \bibnamefont
  {Ramirez}},\ }\bibfield  {title} {\enquote {\bibinfo {title} {Strongly
  geometrically frustrated magnets},}\ }\href {\doibase
  10.1146/annurev.ms.24.080194.002321} {\bibfield  {journal} {\bibinfo
  {journal} {Annual Review of Materials Science}\ }\textbf {\bibinfo {volume}
  {24}},\ \bibinfo {pages} {453--480} (\bibinfo {year} {1994})}\BibitemShut
  {NoStop}%
\bibitem [{\citenamefont {Dyson}(1956)}]{PhysRev.102.1217}%
  \BibitemOpen
  \bibfield  {author} {\bibinfo {author} {\bibfnamefont {Freeman~J.}\
  \bibnamefont {Dyson}},\ }\bibfield  {title} {\enquote {\bibinfo {title}
  {General theory of spin-wave interactions},}\ }\href {\doibase
  10.1103/PhysRev.102.1217} {\bibfield  {journal} {\bibinfo  {journal} {Phys.
  Rev.}\ }\textbf {\bibinfo {volume} {102}},\ \bibinfo {pages} {1217--1230}
  (\bibinfo {year} {1956})}\BibitemShut {NoStop}%
\bibitem [{\citenamefont {Maleev}(1958)}]{Maleev}%
  \BibitemOpen
  \bibfield  {author} {\bibinfo {author} {\bibfnamefont {S.~V.}\ \bibnamefont
  {Maleev}},\ }\href@noop {} {\bibfield  {journal} {\bibinfo  {journal} {Sov.
  Phys. JETP}\ }\textbf {\bibinfo {volume} {64}},\ \bibinfo {pages} {654}
  (\bibinfo {year} {1958})}\BibitemShut {NoStop}%
\bibitem [{\citenamefont {Canali}\ \emph {et~al.}(1992)\citenamefont {Canali},
  \citenamefont {Girvin},\ and\ \citenamefont {Wallin}}]{PhysRevB.45.10131}%
  \BibitemOpen
  \bibfield  {author} {\bibinfo {author} {\bibfnamefont {C.~M.}\ \bibnamefont
  {Canali}}, \bibinfo {author} {\bibfnamefont {S.~M.}\ \bibnamefont {Girvin}},
  \ and\ \bibinfo {author} {\bibfnamefont {Mats}\ \bibnamefont {Wallin}},\
  }\bibfield  {title} {\enquote {\bibinfo {title} {Spin-wave velocity
  renormalization in the two-dimensional heisenberg antiferromagnet at zero
  temperature},}\ }\href {\doibase 10.1103/PhysRevB.45.10131} {\bibfield
  {journal} {\bibinfo  {journal} {Phys. Rev. B}\ }\textbf {\bibinfo {volume}
  {45}},\ \bibinfo {pages} {10131--10134} (\bibinfo {year} {1992})}\BibitemShut
  {NoStop}%
\bibitem [{\citenamefont {Del~Maestro}\ and\ \citenamefont
  {Gingras}(2007)}]{PhysRevB.76.064418}%
  \BibitemOpen
  \bibfield  {author} {\bibinfo {author} {\bibfnamefont {Adrian}\ \bibnamefont
  {Del~Maestro}}\ and\ \bibinfo {author} {\bibfnamefont {Michel J.~P.}\
  \bibnamefont {Gingras}},\ }\bibfield  {title} {\enquote {\bibinfo {title}
  {Low-temperature specific heat and possible gap to magnetic excitations in
  the heisenberg pyrochlore antiferromagnet
  ${\mathrm{gd}}_{2}{\mathrm{sn}}_{2}{\mathrm{o}}_{7}$},}\ }\href {\doibase
  10.1103/PhysRevB.76.064418} {\bibfield  {journal} {\bibinfo  {journal} {Phys.
  Rev. B}\ }\textbf {\bibinfo {volume} {76}},\ \bibinfo {pages} {064418}
  (\bibinfo {year} {2007})}\BibitemShut {NoStop}%
\bibitem [{\citenamefont {Lee}\ and\ \citenamefont {Lee}(2005)}]{Lee05}%
  \BibitemOpen
  \bibfield  {author} {\bibinfo {author} {\bibfnamefont {Sung-Sik}\
  \bibnamefont {Lee}}\ and\ \bibinfo {author} {\bibfnamefont {Patrick~A.}\
  \bibnamefont {Lee}},\ }\bibfield  {title} {\enquote {\bibinfo {title} {U(1)
  gauge theory of the hubbard model: Spin liquid states and possible
  application to
  $\ensuremath{\kappa}\mathrm{\text{-}}(\mathrm{BEDT}\mathrm{\text{-}}\mathrm{TTF}{)}_{2}{\mathrm{cu}}_{2}(\mathrm{CN}{)}_{3}$},}\
  }\href {\doibase 10.1103/PhysRevLett.95.036403} {\bibfield  {journal}
  {\bibinfo  {journal} {Phys. Rev. Lett.}\ }\textbf {\bibinfo {volume} {95}},\
  \bibinfo {pages} {036403} (\bibinfo {year} {2005})}\BibitemShut {NoStop}%
\bibitem [{\citenamefont {Huang}\ \emph {et~al.}(2014)\citenamefont {Huang},
  \citenamefont {Chen},\ and\ \citenamefont {Hermele}}]{huang2014}%
  \BibitemOpen
  \bibfield  {author} {\bibinfo {author} {\bibfnamefont {Yi-Ping}\ \bibnamefont
  {Huang}}, \bibinfo {author} {\bibfnamefont {Gang}\ \bibnamefont {Chen}}, \
  and\ \bibinfo {author} {\bibfnamefont {Michael}\ \bibnamefont {Hermele}},\
  }\bibfield  {title} {\enquote {\bibinfo {title} {Quantum spin ices and
  topological phases from dipolar-octupolar doublets on the pyrochlore
  lattice},}\ }\href {\doibase 10.1103/PhysRevLett.112.167203} {\bibfield
  {journal} {\bibinfo  {journal} {Phys. Rev. Lett.}\ }\textbf {\bibinfo
  {volume} {112}},\ \bibinfo {pages} {167203} (\bibinfo {year}
  {2014})}\BibitemShut {NoStop}%
\bibitem [{\citenamefont {Sibille}\ \emph {et~al.}(2015)\citenamefont
  {Sibille}, \citenamefont {Lhotel}, \citenamefont {Pomjakushin}, \citenamefont
  {Baines}, \citenamefont {Fennell},\ and\ \citenamefont
  {Kenzelmann}}]{PhysRevLett.115.097202}%
  \BibitemOpen
  \bibfield  {author} {\bibinfo {author} {\bibfnamefont {Romain}\ \bibnamefont
  {Sibille}}, \bibinfo {author} {\bibfnamefont {Elsa}\ \bibnamefont {Lhotel}},
  \bibinfo {author} {\bibfnamefont {Vladimir}\ \bibnamefont {Pomjakushin}},
  \bibinfo {author} {\bibfnamefont {Chris}\ \bibnamefont {Baines}}, \bibinfo
  {author} {\bibfnamefont {Tom}\ \bibnamefont {Fennell}}, \ and\ \bibinfo
  {author} {\bibfnamefont {Michel}\ \bibnamefont {Kenzelmann}},\ }\bibfield
  {title} {\enquote {\bibinfo {title} {Candidate quantum spin liquid in the
  ${\mathrm{ce}}^{3+}$ pyrochlore stannate
  ${\mathrm{ce}}_{2}{\mathrm{sn}}_{2}{\mathrm{o}}_{7}$},}\ }\href {\doibase
  10.1103/PhysRevLett.115.097202} {\bibfield  {journal} {\bibinfo  {journal}
  {Phys. Rev. Lett.}\ }\textbf {\bibinfo {volume} {115}},\ \bibinfo {pages}
  {097202} (\bibinfo {year} {2015})}\BibitemShut {NoStop}%
\bibitem [{\citenamefont {Lhotel}\ \emph {et~al.}(2015)\citenamefont {Lhotel},
  \citenamefont {Petit}, \citenamefont {Guitteny}, \citenamefont {Florea},
  \citenamefont {Ciomaga~Hatnean}, \citenamefont {Colin}, \citenamefont
  {Ressouche}, \citenamefont {Lees},\ and\ \citenamefont
  {Balakrishnan}}]{PhysRevLett.115.197202}%
  \BibitemOpen
  \bibfield  {author} {\bibinfo {author} {\bibfnamefont {E.}~\bibnamefont
  {Lhotel}}, \bibinfo {author} {\bibfnamefont {S.}~\bibnamefont {Petit}},
  \bibinfo {author} {\bibfnamefont {S.}~\bibnamefont {Guitteny}}, \bibinfo
  {author} {\bibfnamefont {O.}~\bibnamefont {Florea}}, \bibinfo {author}
  {\bibfnamefont {M.}~\bibnamefont {Ciomaga~Hatnean}}, \bibinfo {author}
  {\bibfnamefont {C.}~\bibnamefont {Colin}}, \bibinfo {author} {\bibfnamefont
  {E.}~\bibnamefont {Ressouche}}, \bibinfo {author} {\bibfnamefont {M.~R.}\
  \bibnamefont {Lees}}, \ and\ \bibinfo {author} {\bibfnamefont
  {G.}~\bibnamefont {Balakrishnan}},\ }\bibfield  {title} {\enquote {\bibinfo
  {title} {Fluctuations and all-in\char21{}all-out ordering in dipole-octupole
  ${\mathrm{nd}}_{2}{\mathrm{zr}}_{2}{\mathrm{o}}_{7}$},}\ }\href {\doibase
  10.1103/PhysRevLett.115.197202} {\bibfield  {journal} {\bibinfo  {journal}
  {Phys. Rev. Lett.}\ }\textbf {\bibinfo {volume} {115}},\ \bibinfo {pages}
  {197202} (\bibinfo {year} {2015})}\BibitemShut {NoStop}%
\bibitem [{\citenamefont {Bertin}\ \emph {et~al.}(2015)\citenamefont {Bertin},
  \citenamefont {Dalmas~de R\'eotier}, \citenamefont {F\aa{}k}, \citenamefont
  {Marin}, \citenamefont {Yaouanc}, \citenamefont {Forget}, \citenamefont
  {Sheptyakov}, \citenamefont {Frick}, \citenamefont {Ritter}, \citenamefont
  {Amato}, \citenamefont {Baines},\ and\ \citenamefont
  {King}}]{PhysRevB.92.144423}%
  \BibitemOpen
  \bibfield  {author} {\bibinfo {author} {\bibfnamefont {A.}~\bibnamefont
  {Bertin}}, \bibinfo {author} {\bibfnamefont {P.}~\bibnamefont {Dalmas~de
  R\'eotier}}, \bibinfo {author} {\bibfnamefont {B.}~\bibnamefont {F\aa{}k}},
  \bibinfo {author} {\bibfnamefont {C.}~\bibnamefont {Marin}}, \bibinfo
  {author} {\bibfnamefont {A.}~\bibnamefont {Yaouanc}}, \bibinfo {author}
  {\bibfnamefont {A.}~\bibnamefont {Forget}}, \bibinfo {author} {\bibfnamefont
  {D.}~\bibnamefont {Sheptyakov}}, \bibinfo {author} {\bibfnamefont
  {B.}~\bibnamefont {Frick}}, \bibinfo {author} {\bibfnamefont
  {C.}~\bibnamefont {Ritter}}, \bibinfo {author} {\bibfnamefont
  {A.}~\bibnamefont {Amato}}, \bibinfo {author} {\bibfnamefont
  {C.}~\bibnamefont {Baines}}, \ and\ \bibinfo {author} {\bibfnamefont
  {P.~J.~C.}\ \bibnamefont {King}},\ }\bibfield  {title} {\enquote {\bibinfo
  {title} {${\mathrm{nd}}_{2}{\mathrm{sn}}_{2}{\mathrm{o}}_{7}$: An
  all-in\char21{}all-out pyrochlore magnet with no divergence-free field and
  anomalously slow paramagnetic spin dynamics},}\ }\href {\doibase
  10.1103/PhysRevB.92.144423} {\bibfield  {journal} {\bibinfo  {journal} {Phys.
  Rev. B}\ }\textbf {\bibinfo {volume} {92}},\ \bibinfo {pages} {144423}
  (\bibinfo {year} {2015})}\BibitemShut {NoStop}%
\bibitem [{\citenamefont {Anand}\ \emph {et~al.}(2015)\citenamefont {Anand},
  \citenamefont {Bera}, \citenamefont {Xu}, \citenamefont {Herrmannsd\"orfer},
  \citenamefont {Ritter},\ and\ \citenamefont {Lake}}]{PhysRevB.92.184418}%
  \BibitemOpen
  \bibfield  {author} {\bibinfo {author} {\bibfnamefont {V.~K.}\ \bibnamefont
  {Anand}}, \bibinfo {author} {\bibfnamefont {A.~K.}\ \bibnamefont {Bera}},
  \bibinfo {author} {\bibfnamefont {J.}~\bibnamefont {Xu}}, \bibinfo {author}
  {\bibfnamefont {T.}~\bibnamefont {Herrmannsd\"orfer}}, \bibinfo {author}
  {\bibfnamefont {C.}~\bibnamefont {Ritter}}, \ and\ \bibinfo {author}
  {\bibfnamefont {B.}~\bibnamefont {Lake}},\ }\bibfield  {title} {\enquote
  {\bibinfo {title} {Observation of long-range magnetic ordering in pyrohafnate
  ${\mathrm{nd}}_{2}{\mathrm{hf}}_{2}{\mathrm{o}}_{7}$: A neutron diffraction
  study},}\ }\href {\doibase 10.1103/PhysRevB.92.184418} {\bibfield  {journal}
  {\bibinfo  {journal} {Phys. Rev. B}\ }\textbf {\bibinfo {volume} {92}},\
  \bibinfo {pages} {184418} (\bibinfo {year} {2015})}\BibitemShut {NoStop}%
\bibitem [{\citenamefont {Xu}\ \emph {et~al.}(2015)\citenamefont {Xu},
  \citenamefont {Anand}, \citenamefont {Bera}, \citenamefont {Frontzek},
  \citenamefont {Abernathy}, \citenamefont {Casati}, \citenamefont
  {Siemensmeyer},\ and\ \citenamefont {Lake}}]{PhysRevB.92.224430}%
  \BibitemOpen
  \bibfield  {author} {\bibinfo {author} {\bibfnamefont {J.}~\bibnamefont
  {Xu}}, \bibinfo {author} {\bibfnamefont {V.~K.}\ \bibnamefont {Anand}},
  \bibinfo {author} {\bibfnamefont {A.~K.}\ \bibnamefont {Bera}}, \bibinfo
  {author} {\bibfnamefont {M.}~\bibnamefont {Frontzek}}, \bibinfo {author}
  {\bibfnamefont {D.~L.}\ \bibnamefont {Abernathy}}, \bibinfo {author}
  {\bibfnamefont {N.}~\bibnamefont {Casati}}, \bibinfo {author} {\bibfnamefont
  {K.}~\bibnamefont {Siemensmeyer}}, \ and\ \bibinfo {author} {\bibfnamefont
  {B.}~\bibnamefont {Lake}},\ }\bibfield  {title} {\enquote {\bibinfo {title}
  {Magnetic structure and crystal-field states of the pyrochlore
  antiferromagnet ${\mathrm{nd}}_{2}{\mathrm{zr}}_{2}{\mathrm{o}}_{7}$},}\
  }\href {\doibase 10.1103/PhysRevB.92.224430} {\bibfield  {journal} {\bibinfo
  {journal} {Phys. Rev. B}\ }\textbf {\bibinfo {volume} {92}},\ \bibinfo
  {pages} {224430} (\bibinfo {year} {2015})}\BibitemShut {NoStop}%
\bibitem [{\citenamefont {Hatnean}\ \emph {et~al.}(2015)\citenamefont
  {Hatnean}, \citenamefont {Lees}, \citenamefont {Petrenko}, \citenamefont
  {Keeble}, \citenamefont {Balakrishnan}, \citenamefont {Gutmann},
  \citenamefont {Klekovkina},\ and\ \citenamefont
  {Malkin}}]{PhysRevB.91.174416}%
  \BibitemOpen
  \bibfield  {author} {\bibinfo {author} {\bibfnamefont {M.~Ciomaga}\
  \bibnamefont {Hatnean}}, \bibinfo {author} {\bibfnamefont {M.~R.}\
  \bibnamefont {Lees}}, \bibinfo {author} {\bibfnamefont {O.~A.}\ \bibnamefont
  {Petrenko}}, \bibinfo {author} {\bibfnamefont {D.~S.}\ \bibnamefont
  {Keeble}}, \bibinfo {author} {\bibfnamefont {G.}~\bibnamefont
  {Balakrishnan}}, \bibinfo {author} {\bibfnamefont {M.~J.}\ \bibnamefont
  {Gutmann}}, \bibinfo {author} {\bibfnamefont {V.~V.}\ \bibnamefont
  {Klekovkina}}, \ and\ \bibinfo {author} {\bibfnamefont {B.~Z.}\ \bibnamefont
  {Malkin}},\ }\bibfield  {title} {\enquote {\bibinfo {title} {Structural and
  magnetic investigations of single-crystalline neodymium zirconate pyrochlore
  ${\mathrm{nd}}_{2}{\mathrm{zr}}_{2}{\mathrm{o}}_{7}$},}\ }\href {\doibase
  10.1103/PhysRevB.91.174416} {\bibfield  {journal} {\bibinfo  {journal} {Phys.
  Rev. B}\ }\textbf {\bibinfo {volume} {91}},\ \bibinfo {pages} {174416}
  (\bibinfo {year} {2015})}\BibitemShut {NoStop}%
\end{thebibliography}%

\end{document}